\documentclass[letterpaper,prl,twocolumn,superscriptaddress,amsmath,amssymb]{revtex4-1}
\setcitestyle{super}
\usepackage{bibunits}
\usepackage{bibentry}
\usepackage{natbib}
\usepackage{graphicx}         
\usepackage{bm}               
\usepackage[colorlinks=true,urlcolor=blue,breaklinks=true]{hyperref}

\newcommand{\linecite}[1]{\hspace{-1 ex} \nocite{#1}\citenum{#1}}

\begin{document}


\begin{bibunit}[naturemag_noURL]
\title{Signatures of interaction-induced helical gaps in nanowire quantum point contacts}

\author{S.~Heedt}
\altaffiliation{Current address: QuTech, Delft University of Technology, 2628 CJ Delft, The Netherlands}
\affiliation{Peter Gr\"unberg Institut (PGI-9) and JARA-Fundamentals of Future Information Technology,
Forschungszentrum J\"ulich, 52425 J\"ulich, Germany}

\author{N.~Traverso Ziani}
\affiliation{Institute of Theoretical Physics and Astrophysics,
University of W\"urzburg, 97074 W\"urzburg, Germany}

\author{F.~Cr\'epin}
\affiliation{Laboratoire de Physique Th\'eorique de la Mati\`ere Condens\'ee, UPMC, CNRS UMR 7600, Sorbonne Universit\'es, 4 place Jussieu, 75252 Paris Cedex 05, France}

\author{W.~Prost}
\affiliation{Solid State Electronics Department, University of Duisburg-Essen, 47057 Duisburg, Germany}

\author{St.~Trellenkamp}
\affiliation{Peter Gr\"unberg Institut (PGI-9) and JARA-Fundamentals of Future Information Technology,
Forschungszentrum J\"ulich, 52425 J\"ulich, Germany}

\author{J.~Schubert}
\affiliation{Peter Gr\"unberg Institut (PGI-9) and JARA-Fundamentals of Future Information Technology,
Forschungszentrum J\"ulich, 52425 J\"ulich, Germany}

\author{D.~Gr\"utzmacher}
\affiliation{Peter Gr\"unberg Institut (PGI-9) and JARA-Fundamentals of Future Information Technology,
Forschungszentrum J\"ulich, 52425 J\"ulich, Germany}

\author{B.~Trauzettel}
\affiliation{Institute of Theoretical Physics and Astrophysics,
University of W\"urzburg, 97074 W\"urzburg, Germany}

\author{Th.~Sch\"apers}
\email{th.schaepers@fz-juelich.de}
\affiliation{Peter Gr\"unberg Institut (PGI-9) and JARA-Fundamentals of Future Information Technology,
Forschungszentrum J\"ulich, 52425 J\"ulich, Germany}

\hyphenation{InAs na-no-wi-re u-sing con-si-der-ing mo-ni-tored na-no-struc-tures na-no-scale spin-tro-nics na-no-elec-tron-ics}

\maketitle

\noindent \textbf{Spin-momentum locking in a semiconductor device with strong spin-orbit coupling (SOC) is a fundamental goal of nanoscale spintronics and an important prerequisite for the formation of Majorana bound states~\cite{Lutchyn2010,Oreg2010,Quay2010}. Such a helical state is predicted in one-dimensional (1D) nanowires subject to strong Rashba SOC and spin-mixing~\cite{Streda2003}, its hallmark being a characteristic reentrant behaviour in the conductance. Here, we report the first direct experimental observations of the reentrant conductance feature, which reveals the formation of a helical liquid, in the lowest 1D subband of an InAs nanowire. Surprisingly, the feature is very prominent also in the absence of magnetic fields. This behaviour suggests that exchange interaction exhibits substantial impact on transport in our device. We attribute the opening of the pseudogap to spin-flipping two-particle backscattering~\cite{Wu2006,Xu2006,Pedder2016}. The all-electric origin of the ideal helical transport bears momentous implications for topological quantum computing.}\\
A 1D conductor with strong SOC is predicted~\cite{Lutchyn2010,Oreg2010,Alicea2011} to represent a viable host for Majorana bound states. These zero-energy states feature characteristic non-Abelian exchange statistics~\cite{Alicea2011} and can be created by mimicking spinless $p$-wave Cooper pairing using a semiconductor nanowire with a helical state and inducing $s$-wave superconductivity. InAs and InSb nanowires are promising host materials to explore the existence and nature of Majorana bound states~\cite{Mourik2012,Das2012}. To this end, it is essential to both establish transport in 1D subbands and induce a helical state in the nanowire. The usual mechanism that is considered to open a helical gap involves an external Zeeman field oriented perpendicular to the uniaxial spin-orbit field~\cite{Streda2003}. The magnitude of the spin-orbit energy relative to the Zeeman energy is partly responsible for the size of the topological energy gap that will protect the zero-energy Majorana modes~\cite{Sau2012}. However, Oreg~\emph{et al.}~\cite{Oreg2010,Oreg2014} and Stoudenmire~\emph{et al.}~\cite{Stoudenmire2011} have pointed out that such an energy gap can also result from strong electronic correlations. Several mechanisms have been proposed along these lines: e.g.\ spin-flipping two-particle backscattering~\cite{Pedder2016} and hyperfine interaction between nuclear spins and a Luttinger liquid~\cite{Braunecker2009}, both of which can open a gap. The latter mechanism has been invoked to explain a conductance reduction by a factor of two at low temperatures in a GaAs quantum wire~\cite{Scheller2014}, but no reentrant behaviour is predicted within this framework.\\
Other than Quay~\emph{et al.}~\cite{Quay2010}, we report on a reentrant conductance feature in the lowest subbands of InAs nanowire quantum point contacts (QPCs), which offer the desired strong SOC (see Supplementary Section~1). Moreover, our proposed spin-mixing mechanism does not necessarily rely on external time-reversal symmetry breaking terms: while the effect is pronounced in the presence of an external magnetic field, it persists also in its absence. Guided by the observation~\cite{Heedt2016} of the Land\'e $g$ factor enhancement for the lowest subband~\cite{Martin2008} and by signatures of the $0.7$ anomaly~\cite{Micolich2011}, we identify the important role of exchange interaction. Following Ref.~\linecite{Pedder2016}, we ascribe the reentrant behaviour at zero magnetic field to the combined presence of 1D confinement, Rashba spin-orbit coupling and Coulomb interaction.
\begin{figure*}[t]
\centering
\includegraphics[width=1.0\linewidth]{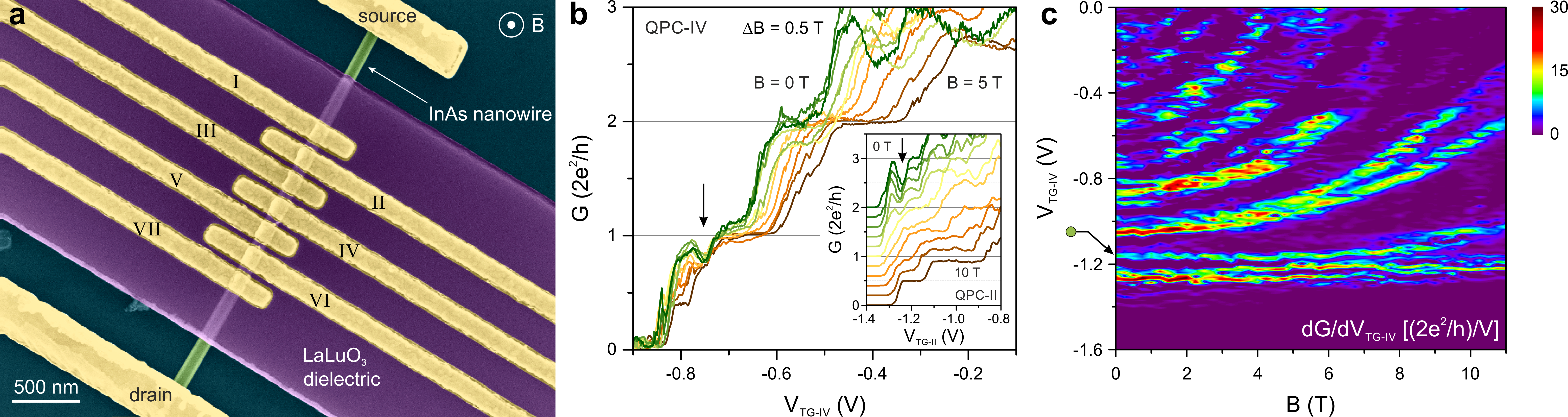}
\caption[device geometry]{\textbf{Quantized conductance and the pseudogap feature.} \textbf{a},~Top-view SEM image of the InAs nanowire covered with a layer of the high-$k$ dielectric LaLuO$_3$. Each of the top-gate electrodes can be used to deplete the channel and control the number of 1D subbands. \textbf{b},~Reentrant conductance behaviour on the first plateau for QPC-IV (dc-bias voltage $V_{\mathrm{dc}}=4\,$mV) at $T=100\,$mK for various magnetic fields $B$. Inset: zero-bias conductance at $T=4\,$K for a different QPC (QPC-II) for $B$ increasing from $0$ to $10\,$T ($\Delta B=1\,$T) with the reentrant conductance feature on the $2\,e^2/h$-plateau. Here, the back-gate voltage is fixed at $V_{\mathrm{BG}}=2\,$V and the curves are offset for clarity. Black arrows indicate the feature location. \textbf{c},~QPC-IV transconductance as a function of $B$ for $V_{\mathrm{dc}}=3\,$mV and $V_{\mathrm{BG}}=2\,$V at $T=3.5\,$K. The high-transconductance line marked by the green dot indicates the evolution of the riser at the edge of the dip feature into the Zeeman-split subband edge at large $B$.}
\label{fig:Sample-Feature}
\end{figure*}~\\
The experimental setup and the main results are summarized in Fig.~\ref{fig:Sample-Feature}. A scanning electron microscopy (SEM) image of the investigated device is depicted in Fig.~\ref{fig:Sample-Feature}a. The contact separation is $2.94\,\mu$m and the wire diameter is $100\,$nm. Top gate fingers of $180\,$nm width can be used to form local QPCs in the nanowire. Enhanced gate coupling is enabled via a high-$k$ dielectric (LaLuO$_3$). The Si/SiO$_2$ back gate can induce an additional electric field in the QPCs. The conductance $G$ in Fig.~\ref{fig:Sample-Feature}b is quantized in integer steps of $2e^2/h$, while at large magnetic field $B$ half-integer steps emerge, reflecting the Zeeman splitting of the first subband, which is determined by a $g$ factor of $\sim7.0$~\cite{Heedt2016}. A single reentrant conductance feature appears reproducibly on the first quantized conductance plateau ($G=2e^2/h$) for magnetic fields smaller than $5\,$T. It exhibits a non-monotonous behaviour, where the conductance drops by up to a factor of two and increases again to the integer quantized conductance value at higher energies, as depicted in Fig.~\ref{fig:Sample-Feature}b and in the inset. The energy spectrum of the QPCs is reflected in the transconductance measurement as a function of the out-of-plane magnetic field. As shown in Fig.~\ref{fig:Sample-Feature}c, the subband edges rise in energy for larger magnetic fields and eventually converge towards Landau levels~\cite{Heedt2016,Heedt2016b}. The riser at the right edge of the reentrant conductance feature is related to the high-transconductance line marked by the green dot in Fig.~\ref{fig:Sample-Feature}c, which continuously evolves into the riser that separates the first two Zeeman-split plateaus at large magnetic fields. This behaviour is expected for a pseudogap that develops into a generic Zeeman splitting.\\
At the single-particle level the physics is rather simple. Owing to the inversion-asymmetry of the device, Rashba SOC gives rise to a spin-dependent shift of the subbands in momentum-space (Fig.~\ref{fig:Helical-Gap_Schematic}a). For an electric field $\mathbf{E}$ perpendicular to the substrate plane, the spin-orbit field $\mathbf{B_{\mathrm{so}}}$ is aligned perpendicular to the nanowire and in the plane of the substrate (cf.\ Fig.~\ref{fig:Helical-Gap_Schematic}d, inset). The spin-orbit energy $E_{\mathrm{so}}=m^*\alpha_{\mathrm{R}}^2/2\hbar^2$ can be calculated from the Rashba coefficient $\alpha_{\mathrm{R}}$ and it represents the energy difference between the degeneracy point at wave number $k=0$ and the band minima, where $m^*$ is the effective electron mass (see Fig.~\ref{fig:Helical-Gap_Schematic}a). In the presence of a strong uniaxial spin-orbit field, a perpendicular magnetic field is expected to open a partial Zeeman gap, giving rise to quasi-helical transport for $g\mu_B B\simeq E_{\mathrm{so}}$~\cite{Streda2003} (cf.\ Fig.~\ref{fig:Helical-Gap_Schematic}b). The formation of a helical state becomes manifest by the appearance of a reentrant conductance plateau at $e^2/h$ inside the larger $2e^2/h$-plateau related to the opening of the pseudogap (Fig.~\ref{fig:Helical-Gap_Schematic}d, top panel). In our experiment, the gap widens roughly proportional to $g\mu_{\mathrm{B}}B$ (see Fig.~\ref{fig:Helical_Gap_TG-IV_vs_B01}). This aspect of the reentrant feature is in accordance with the simple single-particle picture just described. Moreover, with increasing magnetic field, the gap evolves towards the generic Zeeman splitting of a spin-degenerate band (cf.\ Fig.~\ref{fig:Helical-Gap_Schematic}c), which is reflected by the emergence of a plateau at a conductance of $e^2/h$ (Fig.~\ref{fig:Helical-Gap_Schematic}d, bottom panel). However, the experimental observation of the pseudogap feature down to $B=0\,$T (see Fig.~\ref{fig:Helical-Gap_Schematic}e) reveals  the need to go beyond the single-particle picture. The effect we propose to cause the zero-field gap is the combination of Coulomb interaction and the breaking of axial spin symmetry, which can be induced, for instance, by the joint effect of spin-orbit coupling and quantum confinement~\cite{Governale2002}. Since spin is not a conserved quantity, the effective interaction term arising in this framework is correlated two-particle backscattering, which is resonant at $k=0$, similar to the single-particle backscattering caused by the magnetic field. An estimation for the corresponding gap is (see Supplementary Section 2)
\begin{equation}
\Delta_{\mathrm{hel}}=\frac{m^{*4}\left(\alpha_{\mathrm{R}}/\hbar\right)^7e^2d}{\hbar^4\omega_0^3\varepsilon_0\varepsilon_{\mathrm{r}}\sqrt{\hbar/m^*\omega_0}}.
\end{equation}
Using $m^*=0.026\,m_{\mathrm{e}}$ ($m_{\mathrm{e}}$ being the free electron mass), the relative permittivity $\varepsilon_{\mathrm{r}}=15.15$, the confinement energy $\hbar\omega_0=7\,$meV~\cite{Heedt2016} and $\alpha_{\mathrm{R}}=1.2\,$eV$\,$\AA, which is discussed below, only the screening length $d$ remains unknown. Assuming $d=1\,$nm, compatible with the upper bound for the electron density in the QPC segments, and considering the expected exchange-mediated renormalization of the gap by a factor of up to $3$~\cite{Braunecker2009}, we estimate a pseudogap of the order of $\Delta_{\mathrm{hel}}\approx2.5\,$meV, which is compatible with the experimental findings. In fact, using the gate lever arm ($\approx0.04\,$eV/V), we can estimate the energy width of the exchange-mediated gap of $\Delta_{\mathrm{hel}}\approx1.1\,$meV at $T=100\,$mK (e.g.\ see Fig.~\ref{fig:Helical-Gap_Schematic}e).
\begin{figure}[t]
\centering
\includegraphics[width=1.0\linewidth]{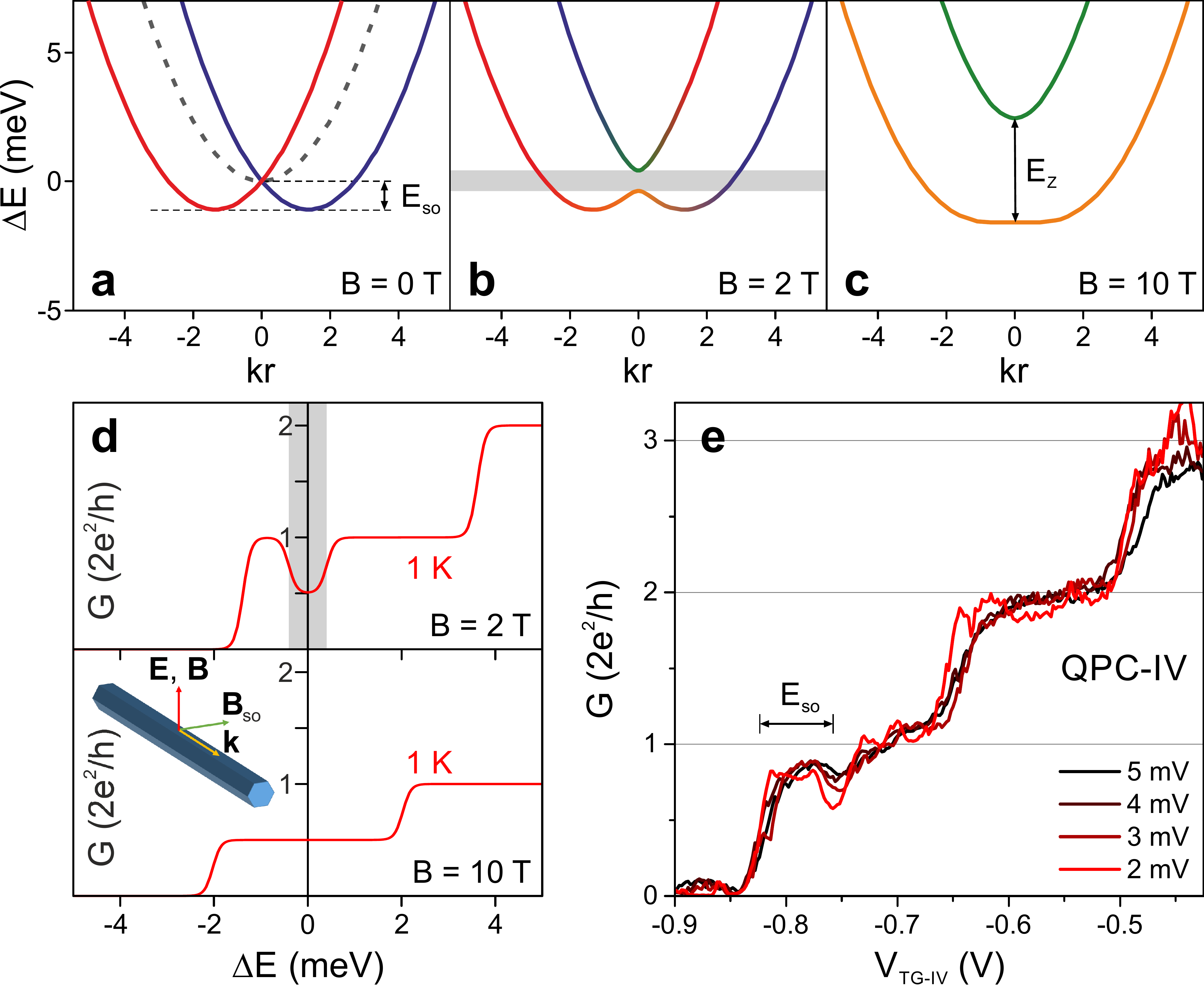}
\caption[helical gap]{\textbf{Helical energy dispersion and bias-dependent reentrant conductance feature.} \textbf{a}--\textbf{c},~Dispersion relations for $\alpha_{\mathrm{R}}=0.8\,$eV$\mathrm{\AA}$ at different magnetic fields $B$ (nanowire radius $r=50\,$nm). \textbf{a},~The two spinful subbands are lowered in energy by $E_{\mathrm{so}}=m^*\alpha_{\mathrm{R}}^2/2\hbar^2$ and shifted in $k$-space by $\pm m^*\alpha_{\mathrm{R}}/\hbar^2$ ($B=0\,$T). \textbf{b},~At $B=2\,$T the two spin bands are mixed by the magnetic field, which is perpendicular to the spin-orbit field $B_{\mathrm{so}}$, leading to an avoided crossing at $k=0$. Red and blue denote spin orientations along the spin-orbit field direction, while green and orange denote spins along the magnetic field direction. \textbf{c},~At $B=10\,$T the Zeeman energy dominates over $E_{\mathrm{so}}$. \textbf{d},~Calculated conductance for the conditions in \textbf{b} (upper panel) and \textbf{c} (lower panel). \textbf{e},~Reentrant feature on the first dc conductance plateau for QPC-IV ($T=100\,$mK) as a function of the dc-bias voltage at $B=0\,$T.}
\label{fig:Helical-Gap_Schematic}
\end{figure}~\\
With regard to the estimate of $\alpha_R$, we observe weak antilocalization in the open, unconfined regime and an avoided crossing in the magnetic field evolution of spin states in quantum dots formed in this device. Both findings substantiate the sizeable SOC in the nanowire (see Supplementary Section 1). Accounting for the gate lever arm, we derive a spin-orbit energy on the order of $2.4\,$meV from the gate voltage position of the centre of the reentrant conductance region (see Fig.~\ref{fig:Helical-Gap_Schematic}e). Thus, $\alpha_{\mathrm{R}}\approx1.2\,$eV$\,$\AA, which is a factor of four larger than $\alpha_{\mathrm{R}}$ derived with conventional methods (cf.\ Supplementary Section~1). It is however of similar magnitude as the Rashba parameters found in InSb nanowires via weak antilocalization, which go up to $1\,$eV$\,$\AA~\cite{vanWeperen2015}. It should be noted that those methods likely underestimate $E_{\mathrm{so}}$, since they only consider spin relaxation in the weakly-confined multi-mode regime.\\
Despite the conceptual simplicity, the visibility of the reentrant behaviour in the conductance for any value of the magnetic field is not to be taken for granted: the unambiguous identification of the SOC-induced conductance feature is generally obstructed by the non-optimal gate potential shape forming the QPC~\cite{Rainis2014} (see Supplementary Section~3). Moreover, the helical gap can be obscured by Fabry-P\'erot resonances that are superimposed on the quantized zero-bias conductance plateaus at low temperatures~\cite{Heedt2016}. Also the shape of the constriction~\cite{Tekman1989} as well as local potential fluctuations~\cite{Nixon1991} can have a critical impact on the transmission. However, at $T>6\,$K the feature can still be observed (see Fig.~\ref{fig:Helical_Gap_TG-IV_vs_B01}a), while Fabry-P\'erot oscillations largely disappear in this regime. This rules out phase-coherent interference as the origin of the effect. At high temperatures ($T\approx9\,$K) the reentrant conductance feature is less pronounced but it broadens with increasing magnetic field and evolves into the first $e^2/h$-plateau for $B>4\,$T (see Fig.~\ref{fig:Helical_Gap_TG-IV_vs_B01}b). Another process that could induce a reentrant behaviour is reflection by impurities. Similar reentrant conductance features have, however, been observed for all investigated QPCs. The reproducibility of the feature position for different QPCs indicates that resonant reflections due to backscattering at impurities in the constrictions~\cite{McEuen1990} are unlikely to explain the observed effect. An additional consistency check is provided by the fact that the reentrant conductance behaviour appears as long as the Zeeman energy is of the order of or smaller than $E_{\mathrm{so}}$, which is a requirement for the existence of the helical gap.
\begin{figure}
\centering
\includegraphics[width=1.0\linewidth]{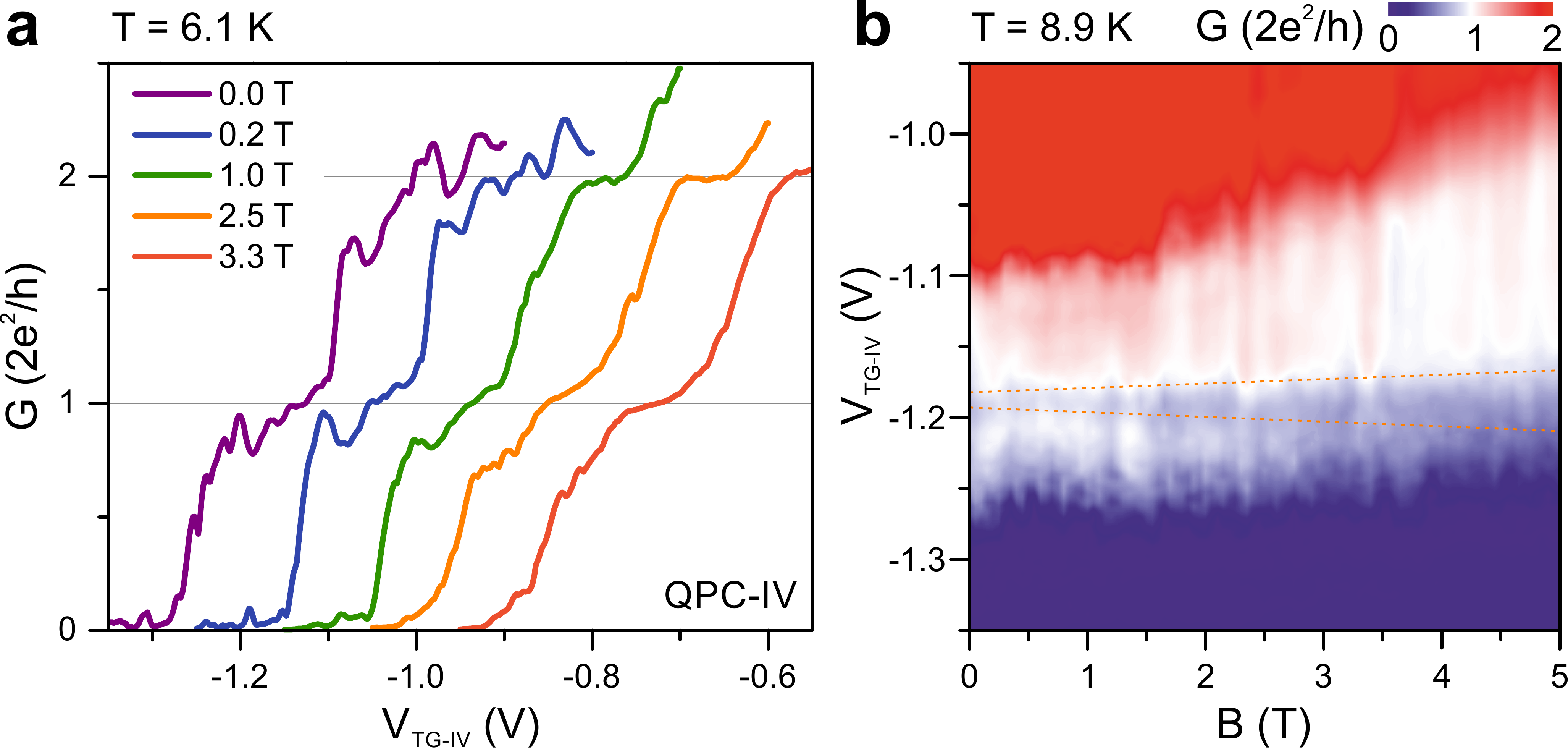}
\caption[helical gap at high T]{\textbf{Reentrant conductance feature at higher temperatures.} Conductance for QPC-IV at (\textbf{a})~$T=6.1\,$K (offset in $V_{\mathrm{TG-IV}}$) and (\textbf{b})~$T=8.9\,$K for different magnetic fields. With increasing magnetic field, the dip feature on the first quantized conductance plateau broadens and the shoulder below the dip disappears at larger $B$. For $B>4\,$T the Zeeman energy contribution dominates over the spin-orbit energy.}
\label{fig:Helical_Gap_TG-IV_vs_B01}
\end{figure}~\\
Further validation of the helical nature of the reentrant behaviour is provided by the analysis of the conductance as a function of Rashba SOC. The magnitude of spin-orbit coupling in the QPC is determined by the strong electric field from the top gate that creates the confinement potential of the constriction. A positive back-gate voltage $V_{\mathrm{BG}}$ does not only increase the conductance of the system, it also enhances the Rashba coefficient at the QPC~\cite{vanWeperen2015}. As depicted in Fig.~\ref{fig:Helical_Gap_TG-IV_vs_Vbg01}, at $B=1.5\,$T the conductance dip is a well-defined singular feature at $V_{\mathrm{BG}}\gg0$. It can be seen that for decreasing $V_{\mathrm{BG}}$ the feature becomes less pronounced and finally develops into a double-plateau, which is characteristic of conventional Zeeman splitting. The modified SOC strength is expected to manifest in a changed visibility of the reentrant conductance feature. For $E_{\mathrm{so}}\ll g\mu_{\mathrm{B}}B$ the shoulder below the dip disappears and a single $e^2/h$-plateau remains.
\begin{figure*}
\centering
\includegraphics[width=1.0\linewidth]{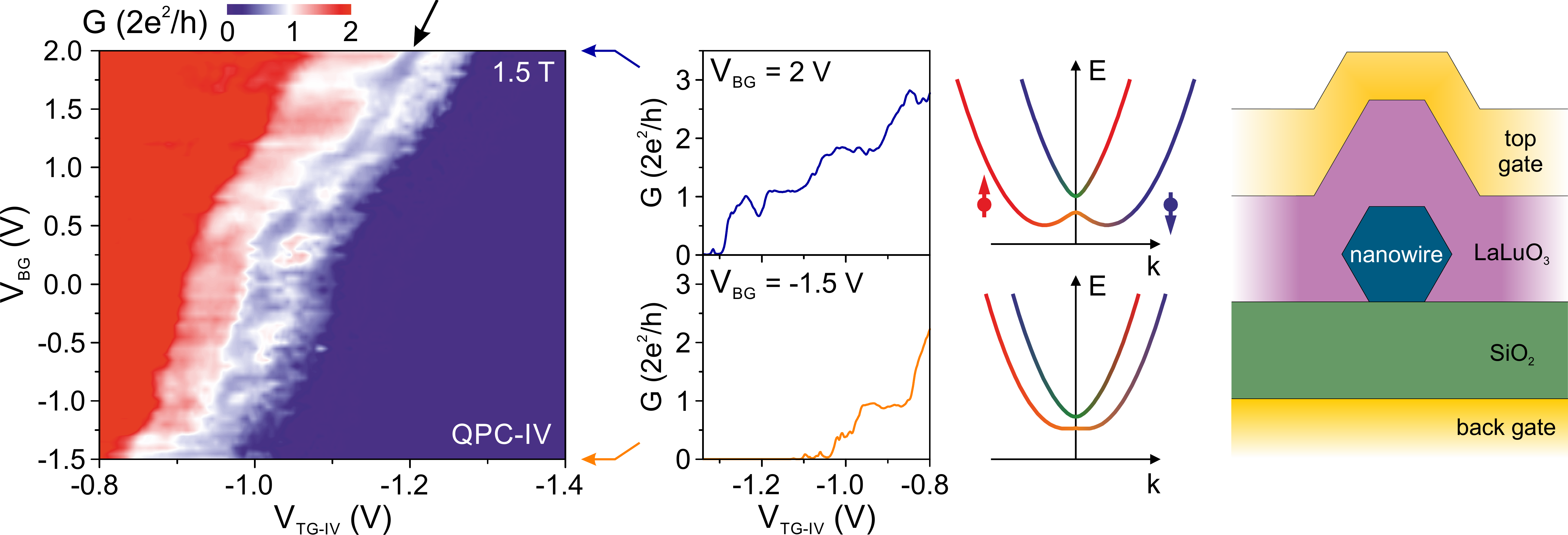}
\caption[helical gap vs Vbg]{\textbf{Impact of the back-gate voltage on the reentrant conductance feature.} Conductance for QPC-IV at $T=4\,$K and $B=1.5\,$T as a function of the back-gate voltage. $V_{\mathrm{BG}}$ changes the Fermi level and also tunes Rashba SOC. The black arrow points at the reentrant feature. The characteristic reentrant behaviour is most pronounced for $V_{\mathrm{BG}}\gg0$, as can be seen in the blue line cut for $V_{\mathrm{BG}}=2\,$V. The corresponding energy dispersion is illustrated to the right of the line cut. The conductance shoulder below the dip feature shrinks for decreasing $V_{\mathrm{BG}}$. For $V_{\mathrm{BG}}\ll0$ the feature disappears and a double-step develops, as depicted in the orange line cut for $V_{\mathrm{BG}}=-1.5\,$V in agreement with the energy dispersion where $E_{\mathrm{Z}}>E_{\mathrm{so}}$. Right: cross-section of the dual-gate device.}
\label{fig:Helical_Gap_TG-IV_vs_Vbg01}
\end{figure*}~\\
In summary, a robust reentrant conductance feature at the $2e^2/h$-plateau is observed for all investigated QPCs along the nanowire. The variation with magnetic field, bias voltage, temperature and back-gate voltage reproduces the signatures of a helical gap. We suggest a spin-mixing mechanism based on strong exchange interaction in order to explain the gap opening at zero magnetic field. The observed spin-orbit energy of $2.4\,$meV is attributed to the strong electric field that generates the distinct confinement in the QPC and enhances Rashba SOC. The all-electric nature of the helical gap preserves time-reversal symmetry and, in the presence of an induced superconducting gap, offers the desired conditions for fractional excitations that give rise to parafermionic quasiparticles~\cite{Pedder2015}, which are promising building blocks in the context of topologically-protected quantum computing~\cite{Hutter2016}.

\section*{Methods}
\noindent \textbf{Device fabrication.} InAs nanowires are grown via gold-catalysed metal-organic vapour phase epitaxy on GaAs (111)B substrates~\cite{Sladek2012}. The field-effect mobility was found to be $25000\,$cm$^2/$Vs~\cite{Heedt2016} and the electron concentration is about $1.0\cdot10^{17}\,$cm$^{-3}$~\cite{Heedt2016b}. Nanowires are mechanically transferred onto a Si substrate with a $200$-nm-thick SiO$_2$ layer that enables back-gate functionality. LaLuO$_3$ dielectric~\cite{Ozben2011} is deposited onto the nanowire via pulsed laser deposition at room temperature and using lift-off technique. The LaLuO$_3$ layer is $100\,$nm thick and the relative permittivity is $\varepsilon_{\mathrm{r}}=26.9$. Ti/Au top-gate electrodes with $180\,$nm width and $30\,$nm pitch are fabricated. Following in situ Ar$^+$ sputtering of the nanowire, $120\,$nm of Ti/Au electrodes are evaporated.\\~\\
\textbf{Measurements.} All measurements have been carried out in a $^3$He/$^4$He dilution refrigerator with a superconducting single-axis magnet. All conductance measurements at zero dc-bias voltage are performed at an ac excitation voltage of $V_{\mathrm{ac}}=80\,\mu$V$_{\mathrm{rms}}$. The bias voltage is applied symmetrically with respect to ground from source to drain electrode and the current is measured simultaneously at both ends of the nanowire. The actual voltage across the QPCs is significantly smaller than the applied voltage considering the voltage drop across the series resistance related to non-ballistic nanowire segments. Typical values of the subtracted series resistance including contributions from the measurement setup are of the order of $20\,$k$\Omega$, chosen such that the conductance plateaus match with integer multiples of $2e^2/h$.
\putbib[MyLibrary]

\section*{Acknowledgments}
\noindent We gratefully acknowledge C.\ Pedder, T.\ Meng and T.\ L.\ Schmidt for fruitful discussions and H. Kertz for valuable help during the measurements. This work was supported by funding from the DFG (SPP1666 and SFB1170 ToCoTronics) and by the Virtual Institute for Topological Insulators (VITI), which is funded by the Helmholtz Association.

\section*{Author contributions}
\noindent S.H., St.T. and J.S. fabricated the sample. S.H. performed the measurements. W.P. grew the InAs nanowires. N.T.Z., F.C. and B.T. derived the anomalous interaction term. S.H., Th.S., N.T.Z., F.C. and B.T. discussed the data. All authors contributed to the manuscript.

\section*{Competing financial interests}
\noindent The authors declare no competing financial interests.
\end{bibunit}

\begin{bibunit}[naturemag_noURL]
\clearpage
\widetext
\begin{center}
\large{\textbf{Supplementary Information: \linebreak Signatures of interaction-induced helical gaps in nanowire quantum point contacts}}\\
\normalsize{~\\
S.~Heedt,$^{1,\ *}$ N.~Traverso Ziani,$^2$ F.~Cr\'epin,$^3$ W.~Prost,$^4$ St.~Trellenkamp,$^1$\\
J.~Schubert,$^1$ D.~Gr\"utzmacher,$^1$ B.~Trauzettel,$^2$ and Th.~Sch\"apers$^{1,\ \dagger}$}\\
\small{~\\
$^1$\textit{Peter Gr\"unberg Institut (PGI-9) and JARA-Fundamentals of Future Information Technology,\\
Forschungszentrum J\"ulich, 52425 J\"ulich, Germany}\\
$^2$\textit{Institute of Theoretical Physics and Astrophysics,\\
University of W\"urzburg, 97074 W\"urzburg, Germany}\\
$^3$\textit{Laboratoire de Physique Th\'eorique de la Mati\`ere Condens\'ee, UPMC,\\
CNRS UMR 7600, Sorbonne Universit\'es, 4 place Jussieu, 75252 Paris Cedex 05, France}\\
$^4$\textit{Solid State Electronics Department, University of Duisburg-Essen, 47057 Duisburg, Germany}}
\end{center}

\setcounter{equation}{0}
\setcounter{figure}{0}
\setcounter{page}{1}
\makeatletter
\renewcommand{\theequation}{S\arabic{equation}}
\renewcommand{\thefigure}{S\arabic{figure}}
\renewcommand{\bibnumfmt}[1]{[S#1]}
\renewcommand{\citenumfont}[1]{S#1}

\section{1. Spin-Orbit Coupling in the Nanowire Device}

\subsection{Weak Antilocalization Effect}
\begin{figure}[h]
\centering
\includegraphics[width=0.5\linewidth]{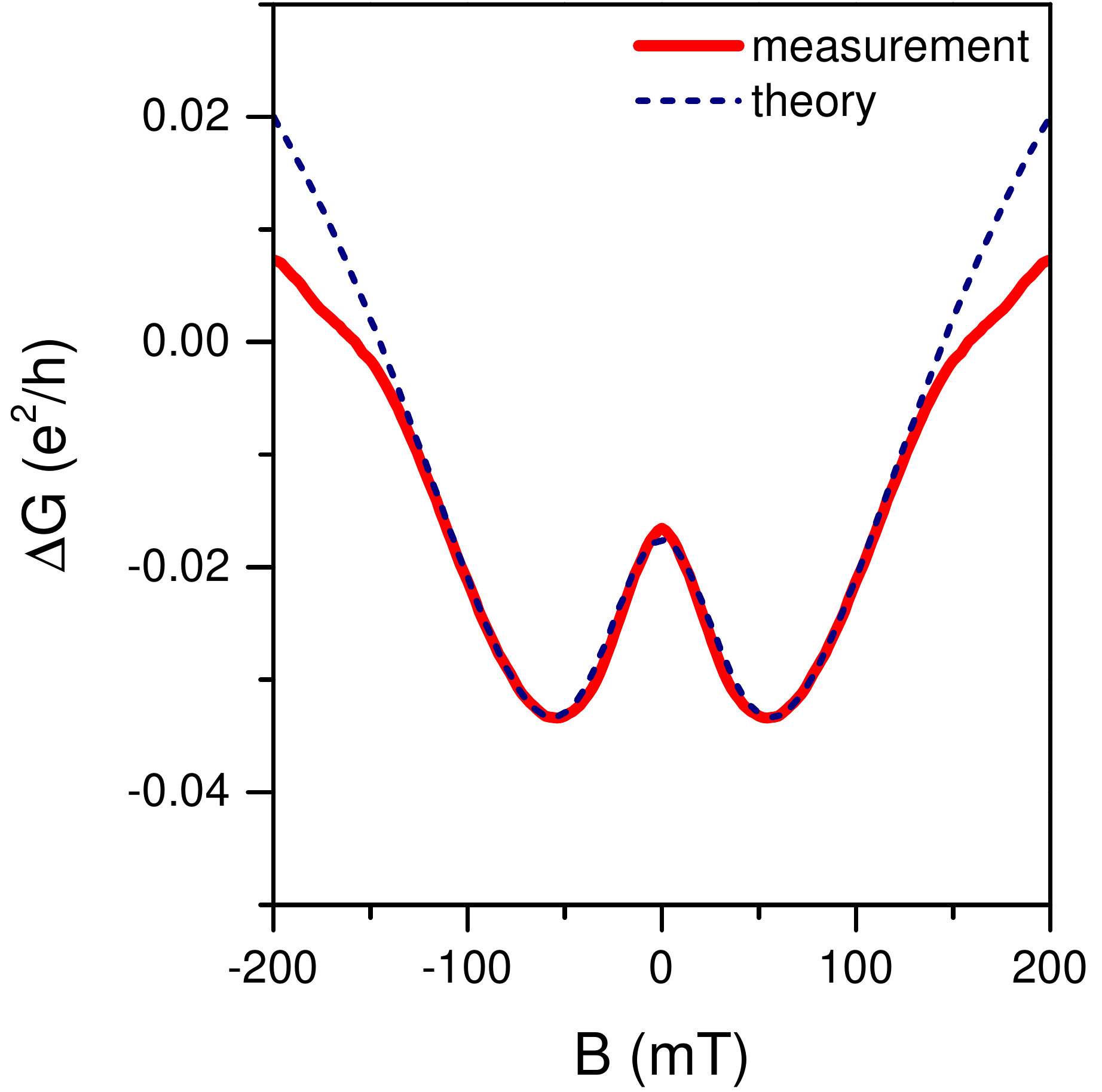}
\caption{\textbf{Weak antilocalization quantum conductance correction.} The sample presented in the main text exhibits a clear signature of the weak antilocalization effect in the low-temperature ($T=50\,$mK) magnetoconductance in the phase-coherent transport regime.}
\label{fig:WAL}
\end{figure}

\noindent The elastic mean free path of $250\,$nm~\cite{Heedt2016} is large for InAs nanowires but the contact separation is more than one order of magnitude larger. Hence, diffusive closed-loop electron trajectories give rise to the weak antilocalization effect in the open, unconfined regime. This is a strong indication for pronounced spin-orbit coupling. It is one of the most common techniques to quantify the relevance of spin-orbit coupling in mesoscopic semiconductors and involves conductance measurements in the phase-coherent transport regime. The signatures of the weak antilocalization effect reveal the presence of spin relaxation that results from spin-orbit coupling~\cite{Hansen2005,vanWeperen2015,Kammermeier2016}. The nanowire conductance at $T=50\,$mK and for a dc-bias voltage of $V_{\mathrm{dc}}=50\,\mu$V is shown in Fig.~\ref{fig:WAL}. Here, $300$ individual magnetoconductance sweeps have been averaged over the back-gate voltage interval from $-1.5\,$V to $1.5\,$V in order to average out magnetoconductance features that are not related to the weak (anti)localization correction to the conductance.\\
The quasiclassical model for the conductance correction that is employed to fit the data is given by~\cite{Beenakker1988,Kurdak1992}
\begin{equation}
\begin{aligned}
\Delta G=-\frac{e^2}{h}\frac{1}{L}&\left[3\left(\frac{1}{l_{\varphi}^2}+\frac{4}{3l_{\mathrm{so}}^2}+\frac{1}{l_B^2}\right)^{-\frac{1}{2}}
-\left(\frac{1}{l_{\varphi}^2}+\frac{1}{l_B^2}\right)^{-\frac{1}{2}}\right.\\
&\left.-3\left(\frac{1}{l_{\varphi}^2}+\frac{4}{3l_{\mathrm{so}}^2}+\frac{d}{l_e^2}+\frac{1}{l_B^2}\right)^{-\frac{1}{2}}
+\left(\frac{1}{l_{\varphi}^2}+\frac{d}{l_e^2}+\frac{1}{l_B^2}\right)^{-\frac{1}{2}}\right],
\end{aligned}
\label{eq:WAL-Kurdak}
\end{equation}
with the magnetic dephasing length
\begin{equation}
l_B=\sqrt{\frac{C_m}{d}\frac{l_m^4}{w^{\gamma_m}l_e^{2-\gamma_m}}}
\label{eq:mag-dephasing}
\end{equation}
and the magnetic length $l_m=\sqrt{\hbar/eB}$. The dimensionality of the unconfined nanowire is $d=3$ since the Fermi wavelength $\lambda_{\mathrm{F}}\approx40\,$nm is smaller than the nanowire diameter ($w=100\,$nm). In the weak magnetic field limit, the magnetic dephasing~\eqref{eq:mag-dephasing} has been quantitatively evaluated by Monte Carlo simulations of the quasiclassical trajectories in a hexagonal nanowire geometry by van Weperen~\emph{et al.}~\cite{vanWeperen2015}. If the magnetic field is aligned perpendicular to the nanowire axis, $C_m=22.3\pm0.3$ and $\gamma_m=3.174\pm0.003$. As depicted in Fig.~\ref{fig:WAL} the fit yields an excellent agreement between the theoretical model and the experimental curve for a phase coherence length $l_{\varphi}=670\pm98\,$nm and a spin relaxation length $l_{\mathrm{so}}=546\pm48\,$nm. Again invoking the Monte Carlo results by van Weperen~\emph{et al.}~\cite{vanWeperen2015}, we can translate the spin relaxation length into a Rashba spin precession length $L_{\mathrm{so}}$ (which is defined as a spin precession by $2\,$rad):
\begin{equation}
l_{\mathrm{so}}=\sqrt{\frac{C_s}{d}\frac{L_{\mathrm{so}}^4}{w^{\gamma_s}l_e^{2-\gamma_s}}}.
\end{equation}
Using the parameters $C_s=8.7\pm0.5$ and $\gamma_s=3.2\pm0.1$~\cite{vanWeperen2015} we obtain $L_{\mathrm{so}}=121\,$nm. Hence, $\alpha_{\mathrm{R}}=\hbar^2/m^*L_{\mathrm{so}}=0.24\,$eV$\,$\AA\ and $E_{\mathrm{so}}=m^*\alpha_{\mathrm{R}}^2/2\hbar^2=100\,\mu$eV. This result is at the upper end of the spin-orbit coupling parameters found in the literature for InAs nanowires. It has to be kept in mind that this value corresponds to the back-gate voltage regime around $V_{\mathrm{BG}}=0\,$V and it has been shown that $\alpha_{\mathrm{R}}$ can increase by a factor of about four upon application of a sizable potential gradient~\cite{Nitta2009,Kammermeier2016}. Such a substantial potential gradient occurs due to the application of a negative top-gate voltage $V_{\mathrm{TG}}$ related to the formation of the local quantum point contacts. Thus, the local top-gate voltage that is used to tune the subband occupation of the quantum point contact has profound impact on the Rashba-type spin-orbit coupling. This holds in particular at the first quantized conductance plateau close to the pinch-off of the 1D channel. The spin-orbit energy $E_{\mathrm{so}}$, which is reflected in the location of the reentrant conductance feature on the $G$-$V_{\mathrm{TG}}$ curve, is actually augmented by the strong electric field of the local gate and the additional contribution from the back gate.

\newpage
\subsection{Singlet-Triplet Anticrossing in a Few-Electron Quantum Dot}
\begin{figure}[h]
\centering
\includegraphics[width=\linewidth]{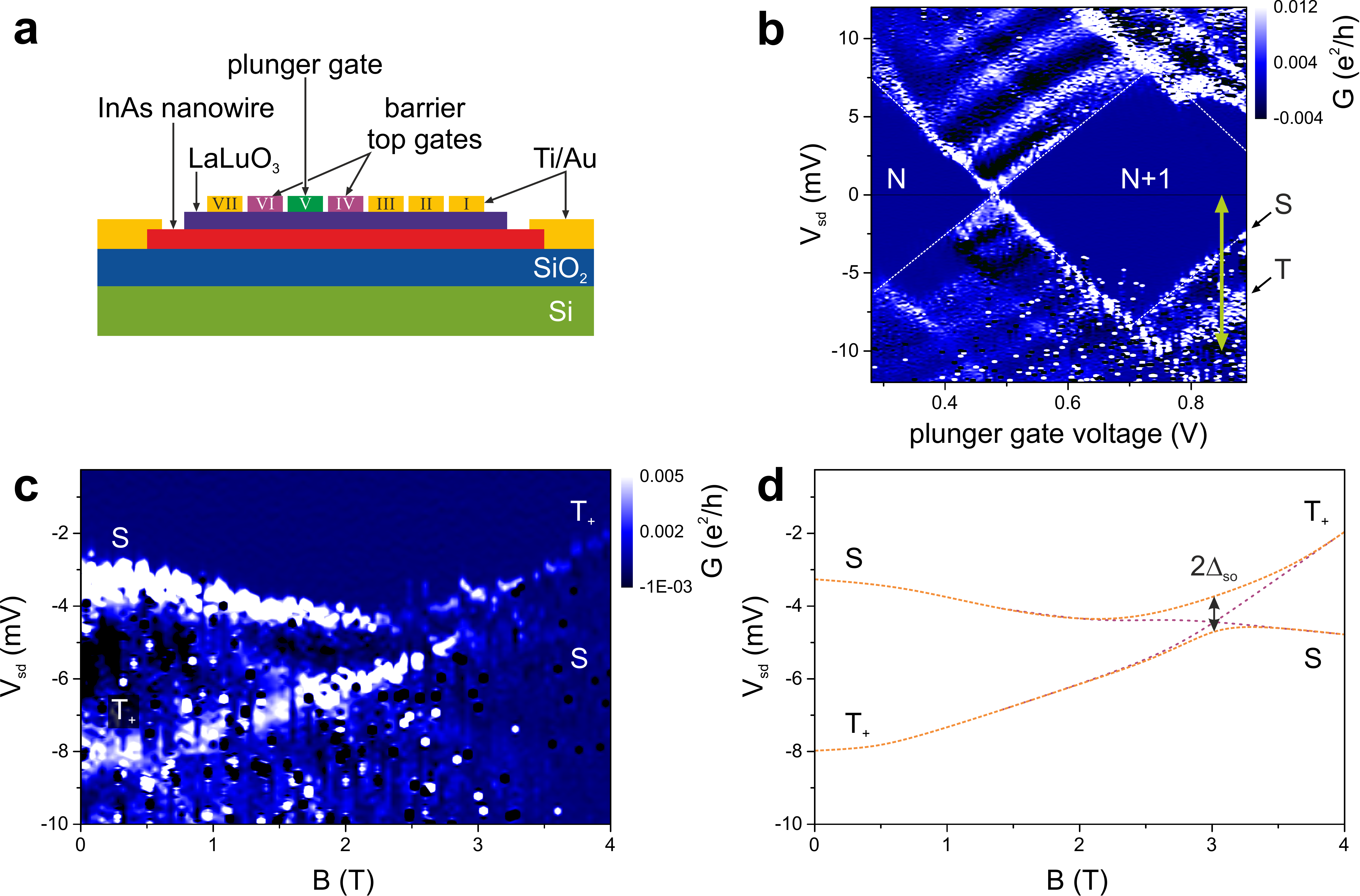}
\caption{\textbf{Sequential transport in a single quantum dot in the few-electron regime.} \textbf{a},~Cross-sectional schematic of the nanowire device. \textbf{b},~Conductance $dI/dV_{\mathrm{sd}}$ as a function of the source-drain voltage $V_{\mathrm{sd}}$ and the plunger gate voltage $V_{\mathrm{TG-V}}$. Top gate~IV ($V_{\mathrm{TG-IV}}=-0.74\,$V) and top gate~VI ($V_{\mathrm{TG-VI}}=-0.76\,$V) are used to create tunnel barriers in the nanowire that define the quantum dot in the device presented in the main text and top gate~V works as a plunger gate which can shift the discrete energy levels of the quantum dot. Outside the Coulomb blockade diamonds the charge stability diagram features a number of lines related to excited states of the quantum dot. The green line indicates the source-drain voltage region which is scanned in (\textbf{c}). \textbf{c},~For an even number of electrons $N$ the highest occupied states are singlets S and triplets T$_{0,\pm}$. Spin-orbit interaction mixes S and T$_+$, which gives rise to an avoided crossing when the states are brought to intersection at $B=3.0\,$T. \textbf{d},~Schematic representation of the avoided crossing between singlet and triplet levels in (\textbf{c}).}
\label{fig:S-T}
\end{figure}

\noindent The top gates can also be employed for an entirely different measurement setup. Although the top-gate width of $180\,$nm is relatively broad, two of the top gates can be used to create two tunnel barriers which enclose a quantum dot, a zero-dimensional charge island. The resulting charging energies are $E_{\mathrm{c}}\simeq6\,$meV and the excitation energies are in the order of $\Delta_{\mathrm{exc}}=2.5\,$meV. We observe Coulomb blockade diamonds in a charge stability diagram and the width of the Coulomb diamonds in terms of plunger gate voltage changes with magnetic field. For odd (even) electron number parity the size of the diamonds is enhanced (reduced) with increasing magnetic field. Electron transport through the quantum dot can be described in terms of sequential tunneling of single charges. If an electron tunnels onto the quantum dot with an odd occupation, it can either occupy a singlet (total spin $S=0$) or a triplet level (total spin $S=1$). Owing to the different spin, the singlet and the triplet state experience a different energy shift in an external magnetic field. Hence, the two states can be brought to intersection, which occurs at $B=3.0\,$T in Fig.~\ref{fig:S-T}. As a signature of spin-orbit coupling, the two states do not intersect but the degeneracy is lifted and an avoided crossing appears. This effect has previously been observed in single quantum dots formed in InAs nanowires~\cite{Fasth2007} and in InSb nanowires~\cite{Nilsson2009}. The magnitude of the avoided crossing is given by $\Delta_{\mathrm{so}}=0.5\,$meV. As an approximation, the avoided crossing can be related to the spin-orbit coupling strength via~\cite{Fasth2007}
\begin{equation}
\Delta_{\mathrm{so}}=\frac{E_{\mathrm{Z}}}{\sqrt{2}}\frac{r_em^*\alpha_{\mathrm{R}}}{\hbar^2},
\label{eq:S-T}
\end{equation}
with the Zeeman energy $E_{\mathrm{Z}}=g\mu_{\mathrm{B}}B$ and the effective electron distance $r_e$.\\
Measuring the Coulomb resonances as a function of the magnetic field yields the $g$ factor of the quantum dot level. We find $g=11$, which is larger compared to the $g$ factors measured in the quantum point contact subbands~\cite{Heedt2016}. This indicates a Zeeman energy at the avoided level crossing of $E_{\mathrm{Z}}=1.9\,$meV. It is well-known that the $g$ factor is strongly diminished due to the orbital confinement~\cite{Kiselev1998}. In our measurement geometry the lateral quantum dot confinement is weak and in the axial direction the quantum dot length is less than $180\,$nm~\cite{Heedt2016b}. Our previous observation that the $g$ factors in the quantum point contacts are significantly reduced compared to the bulk value of $14.7$ indicates that the confinement strength in the quantum point contacts, which are formed right beneath the top gates, is much stronger than in the case of the quantum dots.\\
The effective electron distance can be estimated from the excitation energy, which reflects the axial confinement energy~\cite{Fasth2007}:
\begin{equation}
r_e\approx\sqrt{\hbar^2/m^*\Delta_{\mathrm{exc}}}.
\end{equation}
Thus, $r_e\approx34\,$nm and $\alpha_{\mathrm{R}}\approx0.32\,$eV$\,$\AA. Hence, the spin-orbit energy can be estimated as $E_{\mathrm{so}}=170\,\mu$eV. The Rashba parameter of the order of $\alpha_{\mathrm{R}}\approx0.3\,$eV$\,$\AA\ is in good agreement with the value given above based on the weak antilocalization effect, which is a consistency check supporting the significance of Rashba spin-orbit coupling in our device. As discussed in the main text, the Rashba parameter in the quantum point contacts (QPCs) related to the reentrant conductance feature close to pinch-off is approximately a factor of four larger ($\alpha_{\mathrm{R}}\approx1.2\,$eV$\,$\AA). It is clear that the spin-orbit parameters are qualitatively different since the confinement configuration differs for all three cases. We expect that the confinement is very strong for the case of the QPC, where the constriction forms right under the gate electrode, whereas the quantum dot is formed between two gate electrodes and in the weak antilocalization measurement the top gates are grounded and the channel is in the weakly-confined multi-mode regime.

\newpage~\newpage
\section{2. Two-Particle Backscattering-Induced Helical Gap}
\begin{figure}[h]
\centering
\includegraphics[keepaspectratio, width=1.0\textwidth]{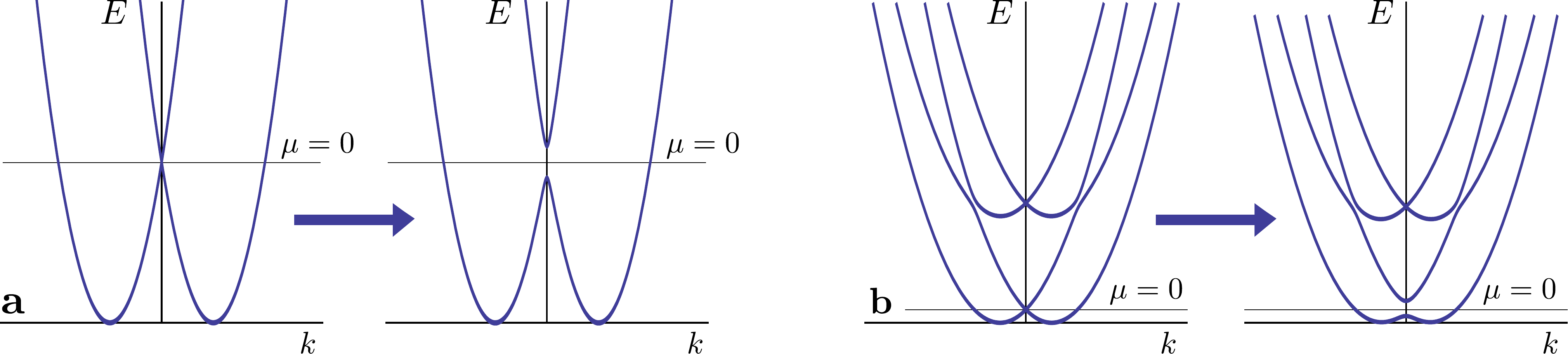}
\caption{\textbf{Different mechanisms responsible for helical gaps in Rashba quantum wires.} \textbf{a} Schematic of the dispersion relation of the first subband of a spin-orbit coupled quantum wire, in case the coupling
with higher subbands is neglected. A gap can be opened at the crossing point at $k=0$ in the presence of a magnetic field applied perpendicular to the spin-orbit field. \textbf{b} Schematic of the dispersion relation of the first two subbands of a spin-orbit coupled quantum wire. An avoided level crossing appears via the hybridization of subbands with different spin and different confinement quantum numbers due to Rashba spin-orbit coupling. Both two-particle backscattering due to Coulomb interactions and applied magnetic fields can open the helical gap at $k=0$.}
\label{fig:1}
\end{figure}

\noindent The Hamiltonian $H_{\mathrm{soc}}$ of a quantum wire with a single occupied confinement subband, in the presence of Rashba spin-orbit coupling and electron-electron interactions, can be conveniently expressed in terms of a Luttinger liquid with a charge and a spin degree of freedom~\cite{Meng2013}. Explicitly,
\begin{equation}
H_{\mathrm{soc}}=\frac{1}{2\pi}\int dx \sum_{\nu=\rho,\sigma}{v_{\nu}}{K_{\nu}}\left(\partial_x \theta_\nu\right)^2+\frac{v_{\nu}}{K_{\nu}}\left(\partial_x \phi_\nu\right)^2.
\label{eq:H}
\end{equation}
Here, $K_\rho<1$ ($K_\sigma>1$), for repulsive interactions, is the Luttinger liquid parameter of the charge (spin) mode, $v_\rho$ ($v_\sigma$) is the corresponding velocity, and $\theta_\nu$ and $\phi_\nu$ are canonically conjugated bosonic fields. The theory in equation~\eqref{eq:H} is gapless. However, it is possible to introduce a gap at momentum $k=0$ and for chemical potential $\mu=0$ (see Fig.~\ref{fig:1}), by applying a magnetic field $B$ perpendicular to the Rashba spin-orbit field. As can be seen in the fermionic picture, the process induced by the magnetic field is a hybridization of the bands at their crossing point by means of a standard Dirac mass term, as shown in Fig.~\ref{fig:1}a. In the bosonization language, the most relevant contribution to the Hamiltonian due to such a magnetic field reads
\begin{equation}
H_B=\frac{g\mu_B B}{2\pi a}\int dx \cos\left[\sqrt{2}\left(\phi_\rho+\theta_\sigma\right)\right],
\end{equation}
where $g$ and $\mu_B$ denote the Land\'e $g$ factor and the Bohr magneton, respectively, and $a$ is the Luttinger liquid cut-off, which can be typically assumed to be of the order of the inverse of the Fermi momentum~\cite{Gindikin2007}. In our case, it is fixed to the spin-orbit wavelength $\hbar^2/m^*\alpha_{\mathrm{R}}$ by the condition $\mu=0$. Similarly, correlated two-particle backscattering
\begin{equation}
H_{\mathrm{2p}}=g_{\mathrm{2p}}\int dx \cos\left[2\sqrt{2}\left(\phi_\rho+\theta_\sigma\right)\right]
\end{equation}
can open a gap at $k=0$ in the absence of external magnetic fields~\cite{Pedder2015,Pedder2016} ($g_{\mathrm{2p}}$ is the coupling constant of two-particle backscattering, and has to be considered, at this stage, a free parameter). In that case, the process amounts to a correlated backscattering of two electrons instead of a single-particle backscattering, as in the case of the magnetic field. However, two ingredients are needed: in order for the contribution $H_{\mathrm{2p}}$ to be relevant in the renormalization group sense, strong electron-electron interactions are essential, so to have $K_\rho+K^{-1}_\sigma<1$, and axial spin symmetry must be broken. A physical mechanism responsible for the breaking of axial spin symmetry could be the coupling among two confinement subbands with different spin projections, which is induced by the Rashba spin-orbit interaction~\cite{Governale2002}, see Fig.~\ref{fig:1}b. In this context, two-particle backscattering emerges after a Schrieffer-Wolff transformation is performed, in order to integrate out the higher subbands and to obtain the effective Hamiltonian for the lowest confinement subband. The helical gap in the second subband should be much smaller than the one in the first subband for two reasons: On the one hand, the Land\'e $g$ factor is smaller and hence, the single-particle effect is weaker. On the other hand, the interaction strength is weaker, most likely, because the particle density is larger. Therefore, it can be expected to observe the helical gap in the first subband but not in the second.
~\\
A detailed description, which leads to the expression for the gap given in the main text once the identification $a\sim \hbar^2/m^*\alpha_{\mathrm{R}}$ is made, is provided in Ref.~\linecite{Pedder2015}. Explicitly, we have started the analysis with the Hamiltonian $H_w$
\begin{equation}
H_w=H_0+H_1+V.
\end{equation}
Here,
\begin{equation}
H_0=\sum_{n,k,s}\epsilon_{n,k,s}c^\dag_{n,k,s}c_{n,k,s},
\end{equation}
where $c_{n,k,s}$ is the fermionic operator for an electron in the subband $n=0,\ldots,\infty$, with wave number $k$ in the $x$ direction and spin $s=\pm 1$. Further, $\epsilon_{n,k,s}=\frac{\hbar^2}{2m^*}(k+\frac{sm^*\alpha_{\mathrm{R}}}{\hbar^2})^2+\hbar\omega n-\frac{m^*\alpha_{\mathrm{R}}^2}{2\hbar^2}$ is the usual kinetic energy term in conjunction with harmonic confinement (typical frequency $\omega$) and the Rashba spin-orbit energy.
\begin{equation}
H_1=-i\alpha_{\mathrm{R}} s\sqrt{\frac{m^*\hbar\omega}{2}}\sqrt{n+1}\left(c^\dag_{n,k,s}c_{n+1,k,-s}-h.c.\right)
\end{equation}
is the relevant subband coupling emerging from the single-particle Rashba term $-\alpha_{\mathrm{R}}\sigma_x p_y$, where $\sigma_x$ is the first Pauli matrix in the usual representation and $p_y$ is the momentum in the confined direction. The electron-electron interaction $V$ reads
\begin{equation}
V=\sum_{n_1,\ldots,n_4}\sum_{s_1,s_2}\sum_{k,k',q}U_{n_1,n_2,n_3,n_4}\left(q\right)c_{n_1,k+q,s}^\dag c_{n_2,k'-q,s'}^\dag c_{n_3,k',s'}c_{n_4,k,s},
\end{equation}
with
\begin{eqnarray}
U_{n_1,n_2,n_3,n_4}\left(q\right)&=&\int \frac{dq_y}{2\pi}U\left(q,q_y\right)\Gamma_{n_1,n_2,n_3,n_4}\left(q_y\right),\\
U\left(q,q_y\right)&=&\int dxdy e^{-i\left(qx+q_yy\right)}U(x,y),\\
\Gamma_{n_1,n_2,n_3,n_4}\left(q_y\right)&=&\int dy_1dy_2 e^{iq_y\left(y_1-y_2\right)}\phi^*_{n_1}\left(y_1\right)\phi^*_{n_2}\left(y_2\right)\phi_{n_3}\left(y_2\right)\phi_{n_4}\left(y_1\right),\nonumber
\end{eqnarray}
where the functions $\phi$ are the eigenfunctions  of the harmonic oscillator.
The explicit expression we have adopted for electron-electron interaction is the screened Coulomb potential $U(x,y)$, given by
\begin{equation}
U\left(x,y\right)=\frac{1}{4\pi\varepsilon_0\varepsilon_r}\left(\frac{1}{\sqrt{x^2+y^2}}-\frac{1}{\sqrt{x^2+y^2+d^2}}\right),
\end{equation}
where $\varepsilon_0$ is the dielectric constant of the vacuum, $\varepsilon_r$ is the dielectric constant relevant for the system, and $d$ is the screening length.
It can be shown that in order to integrate out the higher subbands, a suitable choice for the Schrieffer-Wolff operator $S$ is
\begin{equation}
S=-\left[\sum_{n,k,s}\left(\frac{\sqrt{n+1}\alpha_{\mathrm{R}}\sqrt{\frac{m^*\hbar\omega}{2}}}{\epsilon_{n,k,s}-\epsilon_{n+1,k,-s}}c^\dag_{n+1,k,-s}c_{n,k,s}\right)-h.c.\right].
\end{equation}
By applying the Schrieffer-Wolff transformation to $V$ up to second order, two-particle backscattering terms naturally emerge\cite{Pedder2015,Pedder2016}. Up to a numerical factor of order one, which depends on integrals over the wave functions, their amplitude is the result given in equation~(1) of the main text.\\
Note that the effects of two-particle backscattering and of the applied perpendicular magnetic field are additive. We do not expect any gap closing while increasing the strength of the magnetic field, but rather a monotonic increase of the helical gap.

\newpage
\section{3. Reentrant Conductance Feature}
\begin{figure}[h]
\centering
\includegraphics[width=0.50\linewidth]{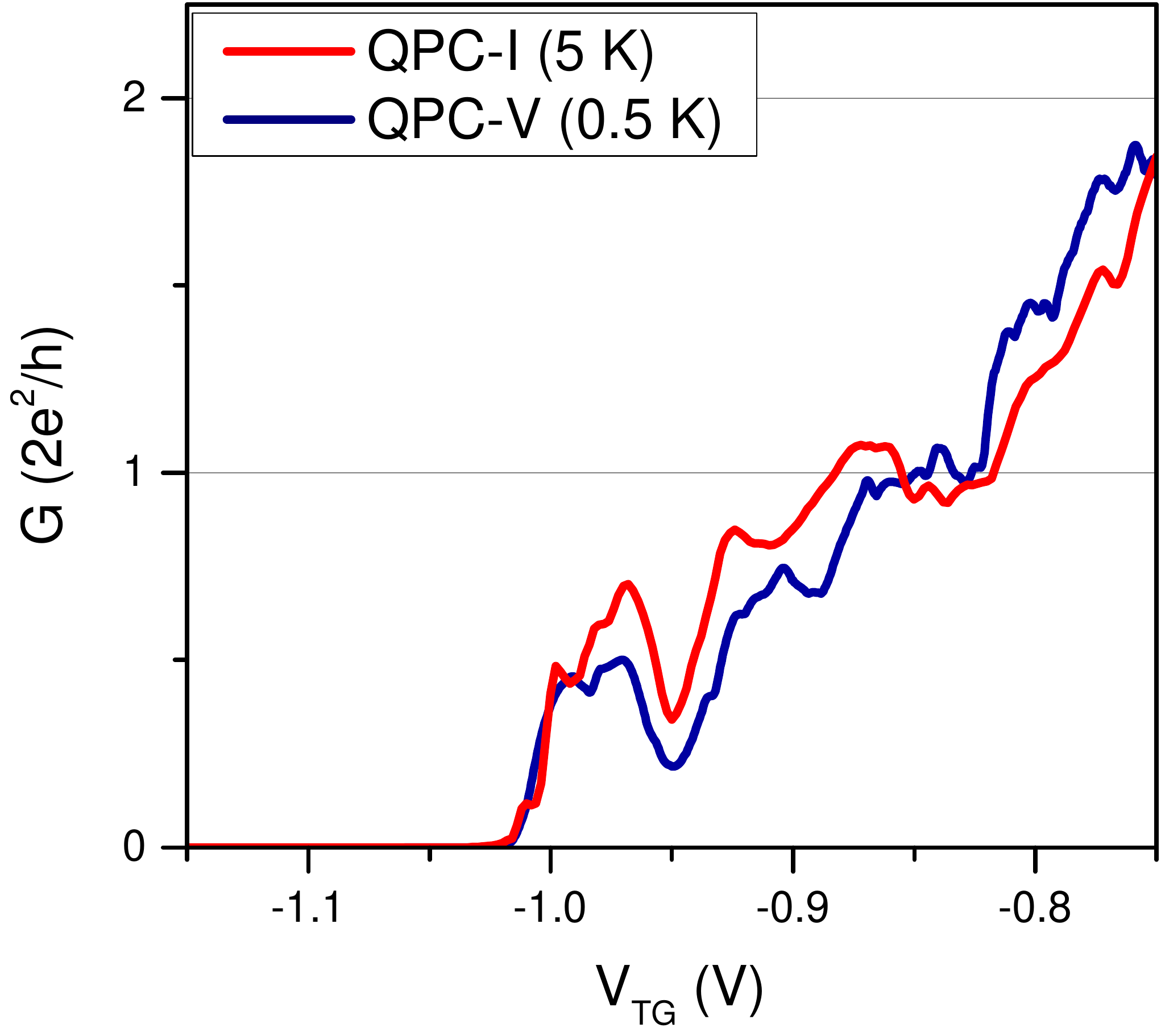}
\caption{\textbf{Signatures of the helical gap for different QPC devices.} The red conductance curve is measured for QPC-I at $T=5\,$K and $V_{\mathrm{dc}}=3\,$mV and the blue curve is measured for QPC-V at a lower temperature of $T=0.5\,$K at a dc-bias voltage of $V_{\mathrm{dc}}=0.8\,$mV for an applied magnetic field of $B=0.4\,$T. The blue curve has been shifted by $-0.53\,$V in top-gate voltage.}
\label{fig:Helical_TGs}
\end{figure}

\begin{figure}[h]
\centering
\includegraphics[width=0.50\linewidth]{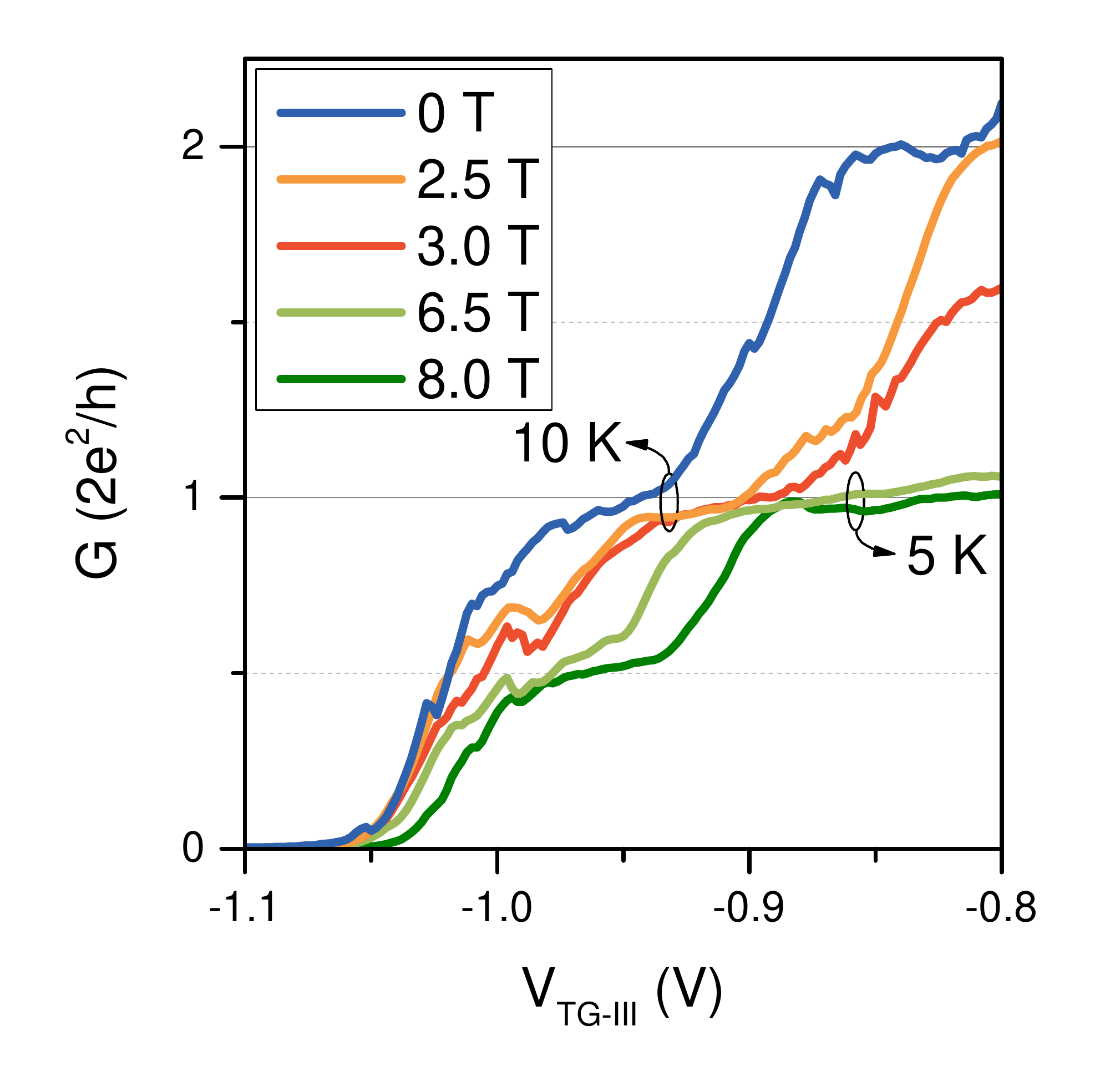}
\caption{\textbf{Reentrant conductance feature at higher temperatures for QPC-III.} Quantized conductance for QPC-III at $T=10\,$K and at zero dc-bias voltage ($V_{\mathrm{ac}}=80\,\mu$V$_{\mathrm{rms}}$). The dip feature on the first quantized conductance plateau appears at $B>0\,$T and evolves into the first half-integer conductance step at $e^2/h$ corresponding to the first non-degenerate spin-up subband. The high-field data (green traces) were taken at $T=5\,$K.}
\label{fig:Helical_TG-III}
\end{figure}

\begin{figure}[h!]
\centering
\includegraphics[width=0.5\linewidth]{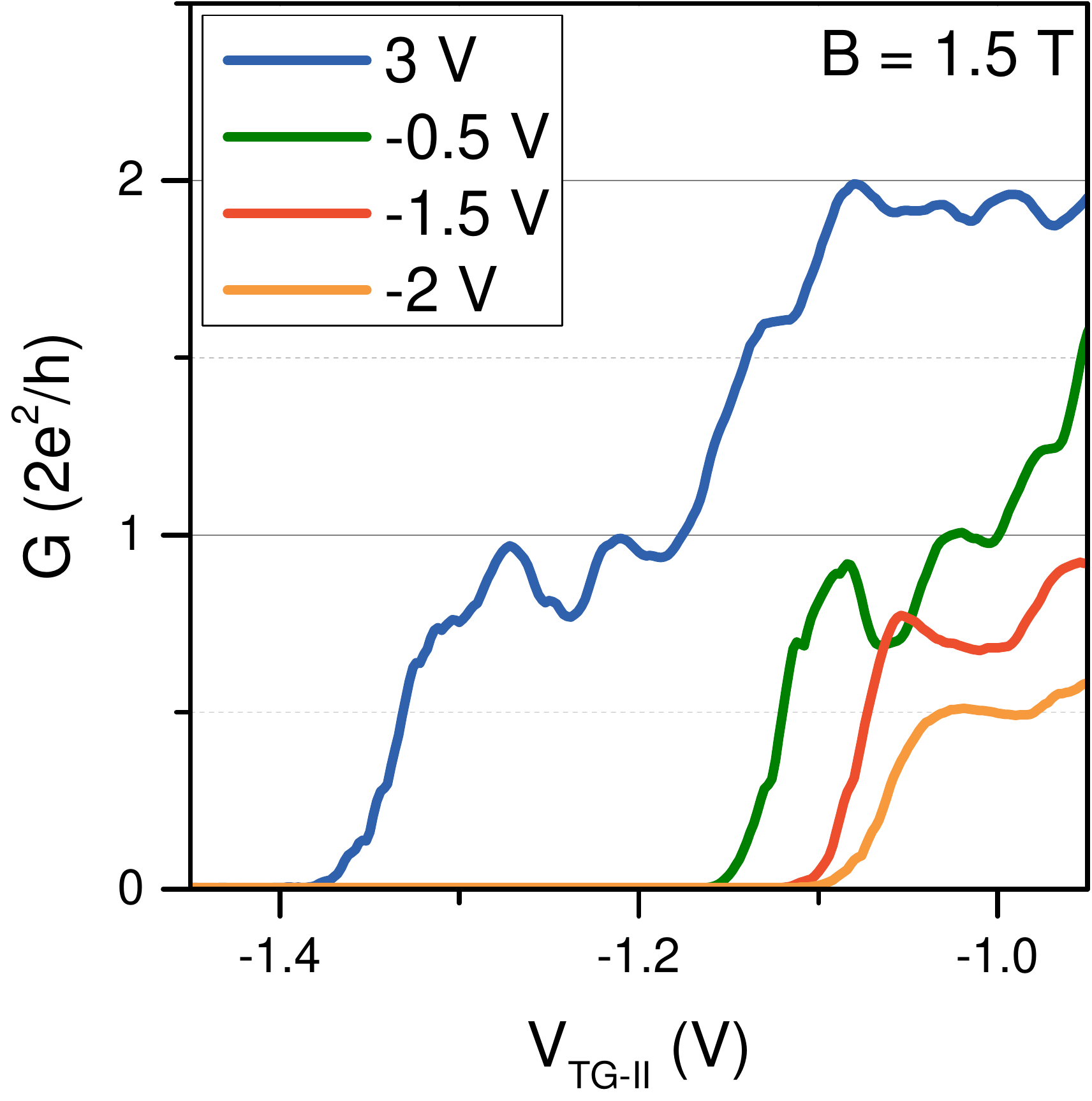}
\caption{\textbf{Impact of the back-gate voltage on the reentrant conductance feature for QPC-II.} Evolution of the reentrant conductance feature in the differential conductance for QPC-II as a function of back-gate voltage. The blue curve has been shifted by $0.1\,$V in top-gate voltage. All measurements are taken at a temperature of $T=4\,$K and for a magnetic field of $B=1.5\,$T.}
\label{fig:Helical_BG-TG-II}
\end{figure}

\noindent All nanowire quantum point contacts that have been investigated have shown a single reentrant feature on the first quantized conductance plateau at $G=2e^2/h$ under certain conditions regarding the temperature, the dc-bias voltage and the back-gate voltage regime. At low temperatures and at small dc-bias voltages, it can be difficult to differentiate the feature from Fabry-P\'erot resonances. However, the feature can also be masked due to bias-voltage or temperature averaging. Usually the reentrant conductance feature is enhanced by applying a positive back-gate voltage. In short constrictions, the conductance in the gap regime rises due to electrons tunneling across the pseudogap region, rendering the reentrant conductance feature imperceptible. The variation of the potential profile creating the QPC occurs on the characteristic length scale $\lambda$ and plays an important role for the adiabaticity criterion~\cite{Rainis2014} $\lambda\sim\lambda^*$, which describes the optimal width of the gate potential profile $\lambda^*=\hbar v_{\mathrm{F}}/(\Delta_{\mathrm{hel}}/2)$. In order to fulfil this visibility condition, a certain ratio between Fermi velocity $v_{\mathrm{F}}$ and helical gap $\Delta_{\mathrm{hel}}$ is required. The optimal QPC length $L$ can be estimated by the width at the half-maximum of the constriction potential $2\lambda^*$. With $v_{\mathrm{F}}$ extracted from the electron concentration ($n=1.0\cdot10^{17}\,$cm$^{-3}$) and $\Delta_{\mathrm{hel}}=1.1\,$meV, $L\approx2\lambda^*=1530\,$nm. Hence, the actual electrostatic dimensions of the QPC potential of $240\,$nm~\cite{Heedt2016b} correspond to the regime $\lambda\approx0.16\lambda^*$, which, according to Rainis and Loss~\cite{Rainis2014}, is close enough to the optimal adiabatic regime to ensure a good visibility of the reentrant region in the conductance.\\
Apart from QPC-II and QPC-IV presented in the main text, also data for QPC-I, III and V are presented below and also exhibit the conductance characteristics related to a partial energy gap. Conductance traces showing a distinct dip feature for QPC-I and QPC-V are depicted in Fig.~\ref{fig:Helical_TGs}. In Fig.~\ref{fig:Helical_TG-III} another dataset is presented for QPC-III for different magnetic field values at zero bias and at high temperatures ($T=10\,$K, except for the two traces at high fields for $B>6\,$T, where $T=5\,$K). There, the blue trace at $B=0\,$T only exhibits a kink attributed to the $0.7$ conductance anomaly~\cite{Heedt2016,Micolich2011} and at intermediate field values a small reentrant conductance feature develops that broadens and eventually evolves into the Zeeman-split $e^2/h$-plateau. This QPC conductance regime is interesting from a fundamental point of view, as  for example Goulko~\emph{et al.}~\cite{Goulko2014} have raised the question of how the $0.7$ anomaly related to strong Coulomb interactions is affected by the presence of pronounced spin-orbit coupling.
~\\
In Fig.~\ref{fig:Helical_BG-TG-II} the impact of the back-gate voltage on the reentrant conductance feature is presented for QPC-II at $B=1.5\,$T, in analogy to Fig.~4 of the main text, where data are presented for QPC-IV. These traces exhibit a corresponding evolution of the conductance from a regime associated with a partial energy gap at positive back-gate voltages towards a double-plateau, which is related to conventional Zeeman splitting. This behaviour is in agreement with the interpretation that the back gate can not only tune the global Fermi level in the nanowire but also the local electric field responsible for Rashba spin-orbit coupling.
\putbib[MyLibrary]
\end{bibunit}


\begin{thebibliography}{10}
\expandafter\ifx\csname url\endcsname\relax
  \def\url#1{\texttt{#1}}\fi
\expandafter\ifx\csname urlprefix\endcsname\relax\def\urlprefix{URL }\fi
\providecommand{\bibinfo}[2]{#2}
\providecommand{\eprint}[2][]{\url{#2}}

\bibitem{Lutchyn2010}
\bibinfo{author}{Lutchyn, R.~M.}, \bibinfo{author}{Sau, J.~D.} \&
  \bibinfo{author}{Das~Sarma, S.}
\newblock \bibinfo{title}{Majorana fermions and a topological phase transition
  in semiconductor-superconductor heterostructures}.
\newblock \emph{\bibinfo{journal}{Phys. Rev. Lett.}}
  \textbf{\bibinfo{volume}{105}}, \bibinfo{pages}{077001}
  (\bibinfo{year}{2010}).

\bibitem{Oreg2010}
\bibinfo{author}{Oreg, Y.}, \bibinfo{author}{Refael, G.} \&
  \bibinfo{author}{von Oppen, F.}
\newblock \bibinfo{title}{Helical liquids and \hbox{Majorana} bound states in
  quantum wires}.
\newblock \emph{\bibinfo{journal}{Phys. Rev. Lett.}}
  \textbf{\bibinfo{volume}{105}}, \bibinfo{pages}{177002}
  (\bibinfo{year}{2010}).

\bibitem{Quay2010}
\bibinfo{author}{Quay, C. H.~L.} \emph{et~al.}
\newblock \bibinfo{title}{Observation of a one-dimensional spin-orbit gap in a
  quantum wire}.
\newblock \emph{\bibinfo{journal}{Nature Phys.}} \textbf{\bibinfo{volume}{6}},
  \bibinfo{pages}{336--339} (\bibinfo{year}{2010}).

\bibitem{Streda2003}
\bibinfo{author}{St\ifmmode~\check{r}\else \v{r}\fi{}eda, P.} \&
  \bibinfo{author}{\ifmmode~\check{S}\else \v{S}\fi{}eba, P.}
\newblock \bibinfo{title}{Antisymmetric spin filtering in one-dimensional
  electron systems with uniform spin-orbit coupling}.
\newblock \emph{\bibinfo{journal}{Phys. Rev. Lett.}}
  \textbf{\bibinfo{volume}{90}}, \bibinfo{pages}{256601}
  (\bibinfo{year}{2003}).

\bibitem{Wu2006}
\bibinfo{author}{Wu, C.}, \bibinfo{author}{Bernevig, B.~A.} \&
  \bibinfo{author}{Zhang, S.-C.}
\newblock \bibinfo{title}{Helical liquid and the edge of quantum spin
  \hbox{Hall} systems}.
\newblock \emph{\bibinfo{journal}{Phys. Rev. Lett.}}
  \textbf{\bibinfo{volume}{96}}, \bibinfo{pages}{106401}
  (\bibinfo{year}{2006}).

\bibitem{Xu2006}
\bibinfo{author}{Xu, C.} \& \bibinfo{author}{Moore, J.~E.}
\newblock \bibinfo{title}{Stability of the quantum spin \mbox{H}all effect:
  Effects of interactions, disorder, and $\mathbb{Z}_2$ topology}.
\newblock \emph{\bibinfo{journal}{Phys. Rev. B}} \textbf{\bibinfo{volume}{73}},
  \bibinfo{pages}{045322} (\bibinfo{year}{2006}).

\bibitem{Pedder2016}
\bibinfo{author}{Pedder, C.~J.}, \bibinfo{author}{Meng, T.},
  \bibinfo{author}{Tiwari, R.~P.} \& \bibinfo{author}{Schmidt, T.~L.}
\newblock \bibinfo{title}{Dynamic response functions and helical gaps in
  interacting \mbox{R}ashba nanowires with and without magnetic fields}.
\newblock \emph{\bibinfo{journal}{Phys. Rev. B}} \textbf{\bibinfo{volume}{94}},
  \bibinfo{pages}{245414} (\bibinfo{year}{2016}).

\bibitem{Alicea2011}
\bibinfo{author}{Alicea, J.}, \bibinfo{author}{Oreg, Y.},
  \bibinfo{author}{Refael, G.}, \bibinfo{author}{von Oppen, F.} \&
  \bibinfo{author}{Fisher, M. P.~A.}
\newblock \bibinfo{title}{Non-\mbox{Abelian} statistics and topological quantum
  information processing in \mbox{1D} wire networks}.
\newblock \emph{\bibinfo{journal}{Nature Phys.}} \textbf{\bibinfo{volume}{7}},
  \bibinfo{pages}{412--417} (\bibinfo{year}{2011}).

\bibitem{Mourik2012}
\bibinfo{author}{Mourik, V.} \emph{et~al.}
\newblock \bibinfo{title}{Signatures of \hbox{Majorana} fermions in hybrid
  superconductor-semiconductor nanowire devices}.
\newblock \emph{\bibinfo{journal}{Science}} \textbf{\bibinfo{volume}{336}},
  \bibinfo{pages}{1003--1007} (\bibinfo{year}{2012}).

\bibitem{Das2012}
\bibinfo{author}{Das, A.} \emph{et~al.}
\newblock \bibinfo{title}{Zero-bias peaks and splitting in an
  \hbox{Al}-\hbox{InAs} nanowire topological superconductor as a signature of
  \hbox{Majorana} fermions}.
\newblock \emph{\bibinfo{journal}{Nature Phys.}} \textbf{\bibinfo{volume}{8}},
  \bibinfo{pages}{887--895} (\bibinfo{year}{2012}).

\bibitem{Sau2012}
\bibinfo{author}{Sau, J.~D.}, \bibinfo{author}{Tewari, S.} \&
  \bibinfo{author}{Das~Sarma, S.}
\newblock \bibinfo{title}{Experimental and materials considerations for the
  topological superconducting state in electron- and hole-doped semiconductors:
  Searching for non-\mbox{Abelian} \mbox{Majorana} modes in \mbox{1D} nanowires
  and \mbox{2D} heterostructures}.
\newblock \emph{\bibinfo{journal}{Phys. Rev. B}} \textbf{\bibinfo{volume}{85}},
  \bibinfo{pages}{064512} (\bibinfo{year}{2012}).

\bibitem{Oreg2014}
\bibinfo{author}{Oreg, Y.}, \bibinfo{author}{Sela, E.} \&
  \bibinfo{author}{Stern, A.}
\newblock \bibinfo{title}{Fractional helical liquids in quantum wires}.
\newblock \emph{\bibinfo{journal}{Phys. Rev. B}} \textbf{\bibinfo{volume}{89}},
  \bibinfo{pages}{115402} (\bibinfo{year}{2014}).

\bibitem{Stoudenmire2011}
\bibinfo{author}{Stoudenmire, E.~M.}, \bibinfo{author}{Alicea, J.},
  \bibinfo{author}{Starykh, O.~A.} \& \bibinfo{author}{Fisher, M.~P.}
\newblock \bibinfo{title}{Interaction effects in topological superconducting
  wires supporting \mbox{M}ajorana fermions}.
\newblock \emph{\bibinfo{journal}{Phys. Rev. B}} \textbf{\bibinfo{volume}{84}},
  \bibinfo{pages}{014503} (\bibinfo{year}{2011}).

\bibitem{Braunecker2009}
\bibinfo{author}{Braunecker, B.}, \bibinfo{author}{Simon, P.} \&
  \bibinfo{author}{Loss, D.}
\newblock \bibinfo{title}{Nuclear magnetism and electron order in interacting
  one-dimensional conductors}.
\newblock \emph{\bibinfo{journal}{Phys. Rev. B}} \textbf{\bibinfo{volume}{80}},
  \bibinfo{pages}{165119} (\bibinfo{year}{2009}).

\bibitem{Scheller2014}
\bibinfo{author}{Scheller, C.~P.} \emph{et~al.}
\newblock \bibinfo{title}{Possible evidence for helical nuclear spin order in
  \mbox{GaAs} quantum wires}.
\newblock \emph{\bibinfo{journal}{Phys. Rev. Lett.}}
  \textbf{\bibinfo{volume}{112}}, \bibinfo{pages}{066801}
  (\bibinfo{year}{2014}).

\bibitem{Heedt2016}
\bibinfo{author}{Heedt, S.}, \bibinfo{author}{Prost, W.},
  \bibinfo{author}{Schubert, J.}, \bibinfo{author}{Gr\"utzmacher, D.} \&
  \bibinfo{author}{Sch\"apers, {\mbox{Th}}.}
\newblock \bibinfo{title}{Ballistic transport and exchange interaction in
  \mbox{InAs} nanowire quantum point contacts}.
\newblock \emph{\bibinfo{journal}{Nano Lett.}} \textbf{\bibinfo{volume}{16}},
  \bibinfo{pages}{3116--3123} (\bibinfo{year}{2016}).

\bibitem{Martin2008}
\bibinfo{author}{Martin, T.~P.} \emph{et~al.}
\newblock \bibinfo{title}{Enhanced \mbox{Z}eeman splitting in
  \mbox{Ga}$_{0.25}$\mbox{In}$_{0.75}$\mbox{As} quantum point contacts}.
\newblock \emph{\bibinfo{journal}{Appl. Phys. Lett.}}
  \textbf{\bibinfo{volume}{93}}, \bibinfo{pages}{012105}
  (\bibinfo{year}{2008}).

\bibitem{Micolich2011}
\bibinfo{author}{Micolich, A.~P.}
\newblock \bibinfo{title}{What lurks below the last plateau: experimental
  studies of the $0.7 \times 2 e^2/h$ conductance anomaly in one-dimensional
  systems}.
\newblock \emph{\bibinfo{journal}{J. Phys. Condens. Matter}}
  \textbf{\bibinfo{volume}{23}}, \bibinfo{pages}{443201}
  (\bibinfo{year}{2011}).

\bibitem{Heedt2016b}
\bibinfo{author}{Heedt, S.} \emph{et~al.}
\newblock \bibinfo{title}{Adiabatic edge channel transport in a nanowire
  quantum point contact register}.
\newblock \emph{\bibinfo{journal}{Nano Lett.}} \textbf{\bibinfo{volume}{16}},
  \bibinfo{pages}{4569--4575} (\bibinfo{year}{2016}).

\bibitem{Governale2002}
\bibinfo{author}{Governale, M.} \& \bibinfo{author}{Z\"ulicke, U.}
\newblock \bibinfo{title}{Spin accumulation in quantum wires with strong
  \mbox{R}ashba spin-orbit coupling}.
\newblock \emph{\bibinfo{journal}{Phys. Rev. B}} \textbf{\bibinfo{volume}{66}},
  \bibinfo{pages}{073311} (\bibinfo{year}{2002}).

\bibitem{vanWeperen2015}
\bibinfo{author}{Van~Weperen, I.} \emph{et~al.}
\newblock \bibinfo{title}{Spin-orbit interaction in \mbox{InSb} nanowires}.
\newblock \emph{\bibinfo{journal}{Phys. Rev. B}} \textbf{\bibinfo{volume}{91}},
  \bibinfo{pages}{201413} (\bibinfo{year}{2015}).

\bibitem{Rainis2014}
\bibinfo{author}{Rainis, D.} \& \bibinfo{author}{Loss, D.}
\newblock \bibinfo{title}{Conductance behavior in nanowires with spin-orbit
  interaction: A numerical study}.
\newblock \emph{\bibinfo{journal}{Phys. Rev. B}} \textbf{\bibinfo{volume}{90}},
  \bibinfo{pages}{235415} (\bibinfo{year}{2014}).

\bibitem{Tekman1989}
\bibinfo{author}{Tekman, E.} \& \bibinfo{author}{Ciraci, S.}
\newblock \bibinfo{title}{Novel features of quantum conduction in a
  constriction}.
\newblock \emph{\bibinfo{journal}{Phys. Rev. B}} \textbf{\bibinfo{volume}{39}},
  \bibinfo{pages}{8772--8775} (\bibinfo{year}{1989}).

\bibitem{Nixon1991}
\bibinfo{author}{Nixon, J.~A.}, \bibinfo{author}{Davies, J.~H.} \&
  \bibinfo{author}{Baranger, H.~U.}
\newblock \bibinfo{title}{Breakdown of quantized conductance in point contacts
  calculated using realistic potentials}.
\newblock \emph{\bibinfo{journal}{Phys. Rev. B}} \textbf{\bibinfo{volume}{43}},
  \bibinfo{pages}{12638--12641} (\bibinfo{year}{1991}).

\bibitem{McEuen1990}
\bibinfo{author}{McEuen, P.~L.}, \bibinfo{author}{Alphenaar, B.~W.},
  \bibinfo{author}{Wheeler, R.~G.} \& \bibinfo{author}{Sacks, R.~N.}
\newblock \bibinfo{title}{Resonant transport effects due to an impurity in a
  narrow constriction}.
\newblock \emph{\bibinfo{journal}{Surf. Sci.}} \textbf{\bibinfo{volume}{229}},
  \bibinfo{pages}{312 -- 315} (\bibinfo{year}{1990}).

\bibitem{Pedder2015}
\bibinfo{author}{Pedder, C.~J.}, \bibinfo{author}{Meng, T.},
  \bibinfo{author}{Tiwari, R.~P.} \& \bibinfo{author}{Schmidt, T.~L.}
\newblock \bibinfo{title}{$\mathbb{Z}_4$ parafermions $\&$ the $8\pi$-periodic
  \mbox{J}osephson effect in interacting \mbox{R}ashba nanowires}.
\newblock \emph{\bibinfo{journal}{ArXiv e-prints}}  (\bibinfo{year}{2015}).
\newblock arXiv:\eprint{1507.08881}.

\bibitem{Hutter2016}
\bibinfo{author}{Hutter, A.} \& \bibinfo{author}{Loss, D.}
\newblock \bibinfo{title}{Quantum computing with parafermions}.
\newblock \emph{\bibinfo{journal}{Phys. Rev. B}} \textbf{\bibinfo{volume}{93}},
  \bibinfo{pages}{125105} (\bibinfo{year}{2016}).

\bibitem{Sladek2012}
\bibinfo{author}{Sladek, K.} \emph{et~al.}
\newblock \bibinfo{title}{Comparison of \hbox{InAs} nanowire conductivity:
  influence of growth method and structure}.
\newblock \emph{\bibinfo{journal}{Phys. Status Solidi C}}
  \textbf{\bibinfo{volume}{9}}, \bibinfo{pages}{230--234}
  (\bibinfo{year}{2012}).

\bibitem{Ozben2011}
\bibinfo{author}{Durgun~{\"O}zben, E.} \emph{et~al.}
\newblock \bibinfo{title}{Integration of \hbox{LaLuO}$_3$ ($\kappa \sim$ 30) as
  high-$\kappa$ dielectric on strained and unstrained \hbox{SOI} \hbox{MOSFET}s
  with a replacement gate process}.
\newblock \emph{\bibinfo{journal}{IEEE Electron Device Lett.}}
  \textbf{\bibinfo{volume}{32}}, \bibinfo{pages}{15--17}
  (\bibinfo{year}{2011}).

\end{thebibliography}


\begin{thebibliography}{10}
\expandafter\ifx\csname url\endcsname\relax
  \def\url#1{\texttt{#1}}\fi
\expandafter\ifx\csname urlprefix\endcsname\relax\def\urlprefix{URL }\fi
\providecommand{\bibinfo}[2]{#2}
\providecommand{\eprint}[2][]{\url{#2}}

\bibitem{Heedt2016}
\bibinfo{author}{Heedt, S.}, \bibinfo{author}{Prost, W.},
  \bibinfo{author}{Schubert, J.}, \bibinfo{author}{Gr\"utzmacher, D.} \&
  \bibinfo{author}{Sch\"apers, {\mbox{Th}}.}
\newblock \bibinfo{title}{Ballistic transport and exchange interaction in
  \mbox{InAs} nanowire quantum point contacts}.
\newblock \emph{\bibinfo{journal}{Nano Lett.}} \textbf{\bibinfo{volume}{16}},
  \bibinfo{pages}{3116--3123} (\bibinfo{year}{2016}).

\bibitem{Hansen2005}
\bibinfo{author}{Hansen, A.~E.}, \bibinfo{author}{Bj\"ork, M.~T.},
  \bibinfo{author}{Fasth, C.}, \bibinfo{author}{Thelander, C.} \&
  \bibinfo{author}{Samuelson, L.}
\newblock \bibinfo{title}{Spin relaxation in \hbox{InAs} nanowires studied by
  tunable weak antilocalization}.
\newblock \emph{\bibinfo{journal}{Phys. Rev. B}} \textbf{\bibinfo{volume}{71}},
  \bibinfo{pages}{205328} (\bibinfo{year}{2005}).

\bibitem{vanWeperen2015}
\bibinfo{author}{Van~Weperen, I.} \emph{et~al.}
\newblock \bibinfo{title}{Spin-orbit interaction in \mbox{InSb} nanowires}.
\newblock \emph{\bibinfo{journal}{Phys. Rev. B}} \textbf{\bibinfo{volume}{91}},
  \bibinfo{pages}{201413} (\bibinfo{year}{2015}).

\bibitem{Kammermeier2016}
\bibinfo{author}{Kammermeier, M.}, \bibinfo{author}{Wenk, P.},
  \bibinfo{author}{Schliemann, J.}, \bibinfo{author}{Heedt, S.} \&
  \bibinfo{author}{Sch{\"a}pers, {\mbox{Th}}.}
\newblock \bibinfo{title}{Weak (anti-)localization in tubular semiconductor
  nanowires with spin-orbit coupling}.
\newblock \emph{\bibinfo{journal}{Phys. Rev. B}}  (\bibinfo{year}{2016}).

\bibitem{Beenakker1988}
\bibinfo{author}{Beenakker, C. W.~J.} \& \bibinfo{author}{van Houten, H.}
\newblock \bibinfo{title}{Boundary scattering and weak localization of
  electrons in a magnetic field}.
\newblock \emph{\bibinfo{journal}{Phys. Rev. B}} \textbf{\bibinfo{volume}{38}},
  \bibinfo{pages}{3232--3240} (\bibinfo{year}{1988}).

\bibitem{Kurdak1992}
\bibinfo{author}{Kurdak, {\mbox{\c{C}}}.}, \bibinfo{author}{Chang, A.~M.},
  \bibinfo{author}{Chin, A.} \& \bibinfo{author}{Chang, T.~Y.}
\newblock \bibinfo{title}{Quantum interference effects and spin-orbit
  interaction in quasi-one-dimensional wires and rings}.
\newblock \emph{\bibinfo{journal}{Phys. Rev. B}} \textbf{\bibinfo{volume}{46}},
  \bibinfo{pages}{6846--6856} (\bibinfo{year}{1992}).

\bibitem{Nitta2009}
\bibinfo{author}{Nitta, J.}, \bibinfo{author}{Bergsten, T.},
  \bibinfo{author}{Kunihashi, Y.} \& \bibinfo{author}{Kohda, M.}
\newblock \bibinfo{title}{Electrical manipulation of spins in the \mbox{R}ashba
  two dimensional electron gas systems}.
\newblock \emph{\bibinfo{journal}{J. Appl. Phys.}}
  \textbf{\bibinfo{volume}{105}} (\bibinfo{year}{2009}).

\bibitem{Fasth2007}
\bibinfo{author}{Fasth, C.}, \bibinfo{author}{Fuhrer, A.},
  \bibinfo{author}{Samuelson, L.}, \bibinfo{author}{Golovach, V.~N.} \&
  \bibinfo{author}{Loss, D.}
\newblock \bibinfo{title}{Direct measurement of the spin-orbit interaction in a
  two-electron \mbox{InAs} nanowire quantum dot}.
\newblock \emph{\bibinfo{journal}{Phys. Rev. Lett.}}
  \textbf{\bibinfo{volume}{98}}, \bibinfo{pages}{266801}
  (\bibinfo{year}{2007}).

\bibitem{Nilsson2009}
\bibinfo{author}{Nilsson, H.~A.} \emph{et~al.}
\newblock \bibinfo{title}{Giant, level-dependent g factors in \mbox{InSb}
  nanowire quantum dots}.
\newblock \emph{\bibinfo{journal}{Nano Lett.}} \textbf{\bibinfo{volume}{9}},
  \bibinfo{pages}{3151--3156} (\bibinfo{year}{2009}).

\bibitem{Kiselev1998}
\bibinfo{author}{Kiselev, A.~A.}, \bibinfo{author}{Ivchenko, E.~L.} \&
  \bibinfo{author}{R\"ossler, U.}
\newblock \bibinfo{title}{Electron g factor in one- and zero-dimensional
  semiconductor nanostructures}.
\newblock \emph{\bibinfo{journal}{Phys. Rev. B}} \textbf{\bibinfo{volume}{58}},
  \bibinfo{pages}{16353--16359} (\bibinfo{year}{1998}).

\bibitem{Heedt2016b}
\bibinfo{author}{Heedt, S.} \emph{et~al.}
\newblock \bibinfo{title}{Adiabatic edge channel transport in a nanowire
  quantum point contact register}.
\newblock \emph{\bibinfo{journal}{Nano Lett.}} \textbf{\bibinfo{volume}{16}},
  \bibinfo{pages}{4569--4575} (\bibinfo{year}{2016}).

\bibitem{Meng2013}
\bibinfo{author}{Meng, T.} \& \bibinfo{author}{Loss, D.}
\newblock \bibinfo{title}{Strongly anisotropic spin response as a signature of
  the helical regime in \mbox{R}ashba nanowires}.
\newblock \emph{\bibinfo{journal}{Phys. Rev. B}} \textbf{\bibinfo{volume}{88}},
  \bibinfo{pages}{035437} (\bibinfo{year}{2013}).

\bibitem{Gindikin2007}
\bibinfo{author}{Gindikin, Y.} \& \bibinfo{author}{Sablikov, V.~A.}
\newblock \bibinfo{title}{Deformed \mbox{W}igner crystal in a one-dimensional
  quantum dot}.
\newblock \emph{\bibinfo{journal}{Phys. Rev. B}} \textbf{\bibinfo{volume}{76}},
  \bibinfo{pages}{045122} (\bibinfo{year}{2007}).

\bibitem{Pedder2015}
\bibinfo{author}{Pedder, C.~J.}, \bibinfo{author}{Meng, T.},
  \bibinfo{author}{Tiwari, R.~P.} \& \bibinfo{author}{Schmidt, T.~L.}
\newblock \bibinfo{title}{$\mathbb{Z}_4$ parafermions $\&$ the $8\pi$-periodic
  \mbox{J}osephson effect in interacting \mbox{R}ashba nanowires}.
\newblock \emph{\bibinfo{journal}{ArXiv e-prints}}  (\bibinfo{year}{2015}).
\newblock arXiv:\eprint{1507.08881}.

\bibitem{Pedder2016}
\bibinfo{author}{Pedder, C.~J.}, \bibinfo{author}{Meng, T.},
  \bibinfo{author}{Tiwari, R.~P.} \& \bibinfo{author}{Schmidt, T.~L.}
\newblock \bibinfo{title}{Dynamic response functions and helical gaps in
  interacting \mbox{R}ashba nanowires with and without magnetic fields}.
\newblock \emph{\bibinfo{journal}{Phys. Rev. B}} \textbf{\bibinfo{volume}{94}},
  \bibinfo{pages}{245414} (\bibinfo{year}{2016}).

\bibitem{Governale2002}
\bibinfo{author}{Governale, M.} \& \bibinfo{author}{Z\"ulicke, U.}
\newblock \bibinfo{title}{Spin accumulation in quantum wires with strong
  \mbox{R}ashba spin-orbit coupling}.
\newblock \emph{\bibinfo{journal}{Phys. Rev. B}} \textbf{\bibinfo{volume}{66}},
  \bibinfo{pages}{073311} (\bibinfo{year}{2002}).

\bibitem{Rainis2014}
\bibinfo{author}{Rainis, D.} \& \bibinfo{author}{Loss, D.}
\newblock \bibinfo{title}{Conductance behavior in nanowires with spin-orbit
  interaction: A numerical study}.
\newblock \emph{\bibinfo{journal}{Phys. Rev. B}} \textbf{\bibinfo{volume}{90}},
  \bibinfo{pages}{235415} (\bibinfo{year}{2014}).

\bibitem{Micolich2011}
\bibinfo{author}{Micolich, A.~P.}
\newblock \bibinfo{title}{What lurks below the last plateau: experimental
  studies of the $0.7 \times 2 e^2/h$ conductance anomaly in one-dimensional
  systems}.
\newblock \emph{\bibinfo{journal}{J. Phys. Condens. Matter}}
  \textbf{\bibinfo{volume}{23}}, \bibinfo{pages}{443201}
  (\bibinfo{year}{2011}).

\bibitem{Goulko2014}
\bibinfo{author}{Goulko, O.}, \bibinfo{author}{Bauer, F.},
  \bibinfo{author}{Heyder, J.} \& \bibinfo{author}{von Delft, J.}
\newblock \bibinfo{title}{Effect of spin-orbit interactions on the 0.7 anomaly
  in quantum point contacts}.
\newblock \emph{\bibinfo{journal}{Phys. Rev. Lett.}}
  \textbf{\bibinfo{volume}{113}}, \bibinfo{pages}{266402}
  (\bibinfo{year}{2014}).

\end{thebibliography}
\end{document}